\documentclass{JHEP3}
\usepackage{epsfig,amsfonts,amsmath}

\newcommand{\fref}[1]{Figure~\ref{#1}}
\newcommand{\cref}[1]{Chapter~\ref{#1}}
\newcommand{\beq}{\begin{equation}}
\newcommand{\eeq}{\end{equation}}
\newcommand{\ba}{\begin{array}}
\newcommand{\ea}{\end{array}}
\newcommand{\bcenter}{\begin{center}}
\newcommand{\ecenter}{\end{center}}

\def\C{\mathbb{C}}

\def\IGa{\relax\hbox{${\rm I}\kern-.18em\Gamma$}}

\def\IT{\mathbb{T}}
\def\Z{\mathbb{Z}}

\def\smiley{\hbox{\large$\bigcirc$\hspace{-0.80em}\raise.2ex
\hbox{$\cdot\cdot$}\kern-.61em\lower.2ex\hbox{\scriptsize$\smile$}}\ }
\def\frowny{\hbox{\large$\bigcirc$\hspace{-0.80em}\raise.2ex
\hbox{$\cdot\cdot$}\kern-.635em\lower.2ex\hbox{\scriptsize$\frown$}}\ }

\makeatletter
\let\hangafter\@hangfrom
\makeatother

\newcommand{\be}{\begin{equation}}
\newcommand{\ee}{\end{equation}}
\newcommand{\bea}{\begin{eqnarray}}
\newcommand{\eea}{\end{eqnarray}}
\newcommand{\bean}{\begin{eqnarray*}}
\newcommand{\eean}{\end{eqnarray*}}
\newcommand{\bc}{\begin{center}}
\newcommand{\ec}{\end{center}}
\newcommand{\comment}[1]{}

\def\kahler{K\"{a}hler\,}

\preprint{MIT-CTP-3702\\ {\tt hep-th/0511063}}

\title{Quivers, Tilings, Branes and Rhombi}

\author{Amihay Hanany$^1$, David Vegh$^1$
\\
~\\
$^1$ Center for Theoretical Physics,
Massachusetts Institute of Technology,\\
Cambridge, MA 02139, USA.\footnote{
Research supported in part by the CTP and the LNS
of MIT and the U.S. Department of Energy under cooperative agreement
$\#$DE-FC02-94ER40818. AH is also supported in part by the BSF American--Israeli Bi--National Science Foundation and 
a DOE OJI award. DV is supported in part by the MIT Praecis Presidential Fellowship.  }\\~\\
\email{hanany@mit.edu, dvegh@mit.edu}
}

\abstract{

We describe a simple algorithm that computes the recently
discovered brane tilings for a given generic toric singular Calabi--Yau threefold.
This therefore gives AdS/CFT dual quiver gauge theories 
for D3--branes probing the given non--compact manifold \cite{Franco:2005rj}.
The algorithm solves a longstanding problem by computing superpotentials
for these theories directly from the toric diagram of the singularity.
We study the parameter space of a--maximization; this study is made possible by identifying
the R--charges of bifundamental fields as angles in the brane tiling.
We also study Seiberg duality from a new perspective.

\comment{
It has been a longstanding open problem to compute the superpotentials for D3--branes 
probing singular non--compact toric Calabi--Yau manifolds.
We describe the ``Fast Inverse Algorithm'' which gives a full solution to this problem.
The algorithm enables one to compute the brane tiling, the quiver graph and the 
superpotential terms from the toric diagram of the singularity.
The R--charges of bifundamental fields are identified as angles in the brane tiling.
This opens the door to study the parameter space of a--maximization and Z--minimization.
We show how fractional branes can be built in the tiling and also make the
connection between zero R--charges and recombining D6--branes.
}

}

\begin{document}

\tableofcontents

\newpage
\section{Introduction}


The AdS/CFT correspondence states that Type IIB string theory on $AdS_5 \times X_5$
is equivalent (dual) to a certain superconformal quiver gauge theory.
Here $X_5$ denotes a five dimensional Sasaki--Einstein\footnote{A
5d manifold is Sasaki--Einstein iff its metric cone is Ricci--flat and \kahler, i.~e. a Calabi--Yau threefold.}
 manifold \cite{Maldacena:1997re, Witten:1998qj, Aharony:1999ti, Klebanov:1998hh,Acharya:1998db,Morrison:1998cs}.
This setup can be constructed by probing a $Y_6$ Calabi--Yau threefold with D3--branes.
The gauge theory emerges on the worldvolume of the branes and its structure reflects
the properties of the singular threefold.
$Y_6$ is the cone over $X_5$ and its metric is related to that on $X_5$ by:
\be
  ds_{Y_6}^2 = dr^2+r^2 ds_{X_5}^2
\ee
The richest structure that is still tractable using the available techniques
can be obtained if we restrict the Calabi--Yau manifold to be toric.
The other four dimensions
are flat, these dimensions are filled by the D3--branes. 
If we set the coordinates of the branes so that they lie at the tip of the cone,
then we obtain the supersymmetric theory on their 3+1 dimensional worldvolume. Its IR limit is then the  field theory dual to the AdS background above.
The Calabi--Yau condition preserves one quarter of the supercharges, the branes further break
half of them, so finally we obtain $\mathcal{N}=1$ supersymmetry in the
four dimensional worldvolume. 
The near horizon limit of this configuration is $AdS_5 \times X_5$.

The matter content of the quiver gauge theory is neatly summarized in the {\bf quiver graph} \cite{Douglas:1996sw}
which also generalizes Dynkin diagrams. Each node in the quiver diagram may carry an index, $N_i$, for the $i$-th node. and denotes $U(N_i)$ gauge group, the edges (arrows) label the
chiral bifundamental fields (see e.~g. \fref{dP0fig1}). These fields transform
in the fundamental representation of $U(N_i)$ and in the anti-fundamental
of $U(N_j)$ where $i$ and $j$ represent the nodes in the quiver that
are the start and endpoints of the corresponding arrow.

The AdS/CFT correspondence is still a conjecture, although it has been justified by many checks.
The comparison of the two dual theories has been hindered by technical difficulties
some of which arise when one is trying to determine the {\bf superpotential} for the quiver gauge theory
which, besides the quiver, one also has to specify for an $\mathcal{N}=1$ supersymmetric theory.

The {\bf Forward Algorithm} \cite{Feng:2000mi, Feng:2001xr} starts with the quiver theory and computes
the toric data for the singularity.
The information contained in the D-- and F--terms can be encoded in a matrix ($Q_t$)
whose cokernel gives the vectors of the toric data (see \cite{Leung:1997tw, fulton} for toric geometry).
The CY condition ($c_1 (Y_6)=0$) implies that these vectors are coplanar, 
so with an appropriate $SL(3,\Z)$ transformation
a convex integer polygon in two dimensions can be obtained.
We will refer to this polygon as the {\bf toric diagram} of the singularity
\cite{Martelli:2004wu, Martelli:2005tp, newVegh}.

\begin{figure}[ht]
  \epsfxsize = 16cm
  \centerline{\epsfbox{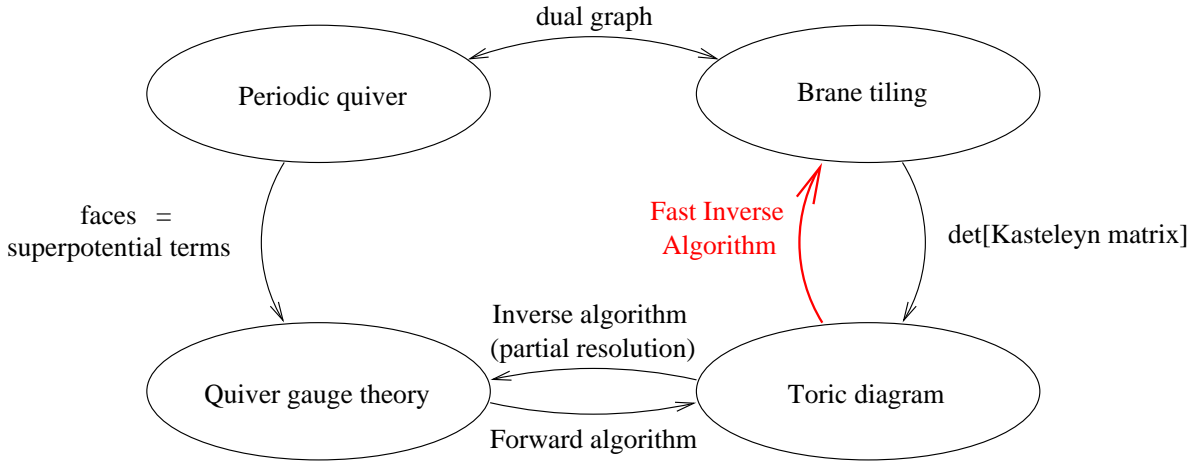}}
  \caption{The logical flowchart.}
  \label{flowchart}
\end{figure}

The brane tiling and the periodic quiver were first introduced in \cite{Franco:2005rj}.
That paper gives a simple algorithm for computing the toric diagram
with multiplicities of the gauged linear sigma model fields using the 
characteristic polynomial of the Kasteleyn matrix in the dimer model 
(see also \cite{Hanany:2005ve, Kenyon:2003uj, Kenyon:2002a}). A logical flowchart between the various concepts is presented in \fref{flowchart}).

In this paper we are going to deal with the following question:
{\bf How does one construct the dual quiver gauge theory for an arbitrary toric singularity?}
We will answer this question by describing the {\bf Fast Inverse Algorithm}
(the red arrow in \fref{flowchart})
which constructs the brane tiling from an arbitrary toric diagram. 
The brane tiling contains
all the information of the quiver gauge theory in the way described above,
so we obtain a method for computing the conformal field theory which is AdS/CFT
dual to the given toric singularity.

A universal method is {\bf partial resolution}
\cite{Douglas:1997de, Beasley:1999uz, Feng:2000mi}.
The Calabi--Yau threefold can be embedded into $\C^3/(\Z_n\times\Z_m)$ where $n$ and $m$
are the smallest integers such that the orbifold toric diagram contains the toric diagram
of the Calabi--Yau manifold of interest.
The dual quiver gauge theory is well--known for abelian orbifolds and we can obtain
the subsector corresponding to the threefold by performing partial resolutions.
The resulting theory is non--unique but flow to the same universality class in the infrared, 
this phenomenon is called toric duality \cite{Feng:2000mi}
which we will investigate in section \ref{section_dualities}.

Another approach utilizes exceptional collections 
of coherent sheafs over divisors which give the quiver gauge theory data \cite{Herzog:2005sy}.
A general algorithm for the computation of tree-level superpotentials was recently 
introduced in \cite{Aspinwall:2004bs, Aspinwall:2005ur} in the context of {\bf derived categories}
\cite{Aspinwall:2004jr}.

In the present paper {\bf we describe a simple algorithm that computes the quiver 
and the superpotential from the toric diagram}.
The next section is devoted to the study of the recently discovered brane tilings 
that will provide us with a very useful tool in constructing the quiver gauge theories.

\newpage

\section{Brane tilings and quivers}
\label{section_branetilings}


\subsection{The basics}
\label{thebasics}

In this section we give a short introduction to the recently discovered brane tilings \cite{Franco:2005rj}.
The {\bf brane tiling} is a configuration of intersecting NS5 and D5--branes in Type IIB string theory that generalize the brane box 
\cite{Hanany:1997tb}
and the brane diamond \cite{Aganagic:1999fe} configurations. This brane configuration has an effective 3+1 dimensional gauge theory on its worldvolume which due to the orientation of the branes has 4 supercharges. This translates to $\mathcal{N}=1$ supersymmetry in four dimensions.

In the brane tiling, there are two types of branes, NS5--brane and D5--branes. The NS5--brane spans the 0123 directions and wraps a holomorphic 
curve in 4567 where the 46 directions are compact.
The D5--branes span the 012346 directions and stretch in between the holes in the network that the NS5--brane forms as it wraps the holomorphic curve. These D5 branes are bounded by the NS5--branes in the 46 directions, 
leading to a 3+1 dimensional theory in their worldvolume at low energies. 
Four supercharges survive the configuration, leading to $\mathcal{N}=1$ supersymmetry in four dimensions. 
This setup is twice T--dual along the 4 and 6 directions to D3--branes probing arbitrary toric singularities.
The NS5--brane is mapped to the singularity
and the D5--branes become probe D3--branes.
This point was demonstrated for the case of brane boxes in \cite{Hanany:1998it} and for the case of brane diamonds in \cite{Aganagic:1999fe}. Furthermore for brane intervals of the type introduced in \cite{Hanany:1996ie} T--duality needs to be done once to get to a configuration of D3 branes probing a singular CY manifold and the relation between the two constructions is studied in \cite{Karch:1998yv}
where the dualities act on the compact 46 directions.
As we will soon show, the brane tiling graph encodes the quiver and the superpotential information,
therefore fully specifies the 4D $\mathcal{N}=1$ theory.

\begin{figure}[ht]
\begin{center}
  \epsfxsize = 4cm
  \centerline{\epsfbox{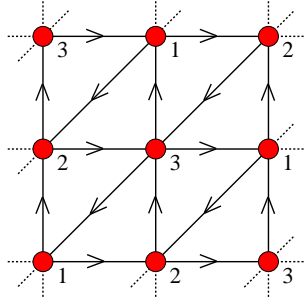}}
  \caption{The $\bf{dP}_0$ periodic quiver. The nodes denote $U(N)$ gauge groups,
the directed edges between them are bifundamental fields. The plaquettes of the
quiver graph are terms in the superpotential. This example has three gauge groups,
they are labelled by numbers. If we identify the nodes with the same numbers (i.~e. we ``compactify''
the periodic quiver), then we arrive at the usual quiver diagram.}
  \label{infq}
\end{center}
\end{figure}

The relevant physics is visualized by drawing the brane tiling in the 46 plane.
The intersection of the holomorphic curve in 4567 with this plane is the brane tiling graph.
This tiling is doubly periodic since the 46 directions are taken 
to be compact. The orientation of the NS5--brane together with the holomorphicity of the curve it wraps implies that the graph in the 46 directions is bipartite.\footnote{A graph is bipartite when 
its nodes can be colored in white and black, such that 
edges only connect black nodes to white nodes and vice versa. This implies that each face 
has even number of edges.} This bipartite property is identified with the orientation of fundamental strings along an intersection point of NS5--branes. \fref{coniffig} shows an example of this property. The green arrows indicate orientations of fundamental strings stretching between two neighboring D5 branes and around a white node these are oriented in a clockwise fashion. On the other hand, around a black node the strings are oriented in a counterclockwise fashion. The bipartite property implies that each face in the brane tiling has an even number of edges and that it has equal number of incoming and outgoing arrows. This implies anomaly cancellation in the quiver gauge theory which ensures that this gauge theory is well--defined.
Another useful concept which follows from the brane tiling is its dual graph which is termed the {\bf periodic quiver}. The periodic quiver is a special type of quiver, with two periodic directions in which nodes and arrows are identified across two directions. \fref{infq} shows an example of a periodic quiver for the well known case of $\C^3/\Z_3$ (otherwise known as the complex cone over $\bf{dP}_0$). Nodes carry 3 different labels and nodes with the same label as well as arrows between them are identified.

\be
\begin{array}{cc|cc|ccc}
 \mbox{{\bf Brane tiling}} & & \
 \mbox{{\bf Periodic quiver}} & &\ \ & \mbox{{\bf Gauge theory}} \\ 
 \hline \hline
 \mbox{faces} & & \mbox{nodes} & & & U(N) \ \mbox{gauge groups} \\
 \mbox{edges} & & \mbox{edges} & & & \mbox{bifundamental fields} \\
 \mbox{nodes} & & \mbox{plaquettes} & & & \mbox{superpotential terms} 
\end{array}
\nonumber
\label{dictfig}
\ee

Given a brane tiling, it is straightforward to derive the associated quiver gauge theory. 
The tiling encodes both the quiver diagram and the superpotential, which can be constructed 
in the following way. The dual graph to the tiling is the periodic quiver (see e.~g. \fref{infq}).
The periodic quiver can be seen as the usual quiver graph drawn on the surface of a 2--torus.
The plaquettes of the periodic quiver
are the terms in the superpotential. These plaquettes correspond to black and white nodes 
in the brane tiling. The color of the node in the tiling tells us the sign,
the valence of the node is equal to the order of the term.
Thus, we conclude that each bifundamental field appears exactly twice in the superpotential, 
once with a plus and once with a minus sign.
We see that the tiling provides us with a simple geometrical unification of quiver 
and superpotential data.

\begin{figure}[ht]
\begin{center}
  \epsfxsize = 11cm
  \centerline{\epsfbox{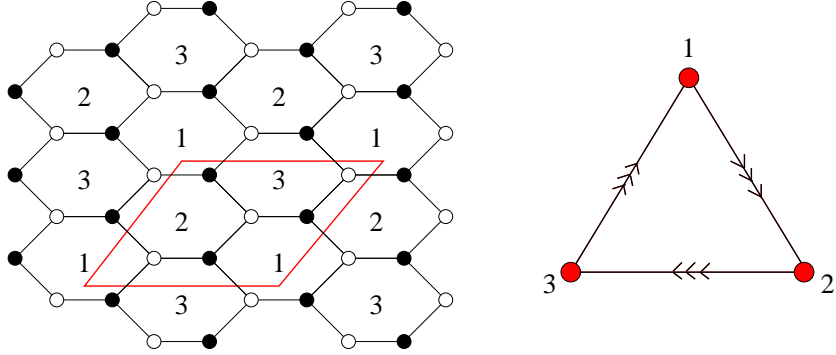}}
  \caption{$\bf{dP}_0$ brane tiling \& quiver. The unit cell of the lattice is shown in red.
The theory has three gauge groups (faces in the tiling) and six cubic terms in the superpotential
(valence three nodes of the tiling).}
  \label{dP0fig1}
\end{center}
\end{figure}

As an example, \fref{dP0fig1} shows the brane tiling and the quiver for $\bf{dP}_0$.
We see that the brane tiling contains three faces, these correspond to the three
gauge groupes (nodes) in the quiver. The nine edges in the tiling are the bifundamental fields.
The six nodes of the tiling immediately give the following superpotential:
\bea
\nonumber
  W=X_{12}^{(1)} X_{23}^{(2)} X_{31}^{(3)} + X_{12}^{(2)} X_{23}^{(3)} X_{31}^{(1)}+X_{12}^{(3)} X_{23}^{(1)} X_{31}^{(2)} \\
-X_{12}^{(3)} X_{23}^{(2)} X_{31}^{(1)} - X_{12}^{(2)} X_{23}^{(1)} X_{31}^{(3)}  - X_{12}^{(1)} X_{23}^{(3)} X_{31}^{(2)}
\eea
Here $X_{ij}^{(k)}$ denotes the bifundamentals going from gauge group $i$ to $j$, and $k$ is
just labelling the different fields.

Another example is the conifold (see Figure 4). The tiling contains two faces
which correspond to the two gauge groups. 

\begin{figure}[ht]
  \epsfxsize = 10cm
  \centerline{\epsfbox{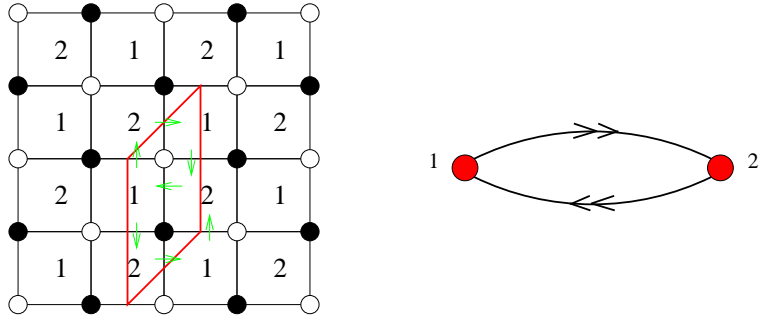}}
  \caption{Brane tiling \& quiver for the conifold. The green arrows indicate the directions of the bifundamental arrows in the quiver}
\label{coniffig}
\end{figure}

From the tiling we can read off the two quadratic terms in the superpotential:
\be
  W=X_{12}^{(1)} X_{21}^{(1)} X_{12}^{(2)} X_{21}^{(2)} -X_{12}^{(1)} X_{21}^{(2)} X_{12}^{(2)} X_{21}^{(1)}
\ee

In general, the number of gauge groups in the quiver, or equivalently, the number of faces in the 
brane tiling is equal to twice the area of the toric diagram \cite{Feng:2002zw}.
The area of an arbitrary integer polygon can be calculated 
by means of {\bf Pick's theorem} \cite{Pick}:
\be
2 \cdot \mbox{Area} =  2I+E-2,
\label{Pick}
\ee
where $I$ is the number of internal points in the toric diagram , and $E$ denotes the number of external points (points on the edges) in the toric diagram.

Using this formula we can relate the area of the toric diagram to the number of gauge groups in the quiver gauge theory. The number of 0-cycles, 2-cycles and 4-cycles in the non-compact CY manifold are given by $1$, $I+E-3$ and $I$ respectively. Using the relation that the number of gauge groups is equal to the sum of these three numbers we arrive at the relation that the number of gauge groups is twice the area.

For a complete introduction to brane tilings and to the Fast Forward Algorithm and for more examples
the reader should refer to \cite{Franco:2005rj, Hanany:2005ve}.

\subsection{Superconformal fixed point and R--charges}
\label{section_amaxi}

In the last section we reviewed the construction of brane tiling. 
In this section we are going to find a new connection between R--charges 
of fundamental fields and some basic properties of the tiling configuration.

The quiver gauge theories described by the brane tilings are expected to flow 
at low energies to a superconformal fixed point.
The global symmetry group of the theory contains the $U(1)$ R--symmetry.
The Sasaki--Einstein manifolds have a canonical Killing vector field called the Reeb vector.
This is dual to the R--symmetry of the quiver gauge theory.

It has been shown in \cite{Intriligator:2003jj} that the superconformal R--charges 
can be determined by a--maximization: the R--symmetry is the  $U(1)$ symmetry, 
which maximizes the combination of 't~Hooft anomalies 
$a(R) \equiv (9 Tr R^3-3 Tr R)/32$. 
The maximal value of $a$ is then suggested to be the central charge of the superconformal theory
(for details see \cite{Anselmi:1997am, Anselmi:1997ys}).

The R--charges are related by the AdS/CFT correspondence to volumes of
supersymmetric submanifolds in the dual Sasaki--Einstein manifold.
Recently, it has been shown \cite{Martelli:2005tp} that
these volumes can be extracted from the toric data 
of the Calabi--Yau singularity without knowing the metric explicitly.
The R--charges can be obtained by minimizing a function $Z$ 
that depends only on the toric data of the singularity and the trial Reeb vector.
This method is called the geometric dual to a--maximization.


Let us assign an R--charge to each bifundamental field in the brane tiling. 
At the IR superconformal fixed point, each term in the superpotential 
satisfies
\beq
\sum_{ i \in edges \,\, around \,\, node } R_i =2 \qquad {\rm for\,\, each \,\,node}
\label{wcharge}
\eeq
where the sum is over all edges surrounding a given node.
The (numerator of the) NSVZ beta function for each gauge coupling vanishes, which leads to the following equation:
\beq
\label{wch20}
\sum_{i \in edges\,\, around \,\, face} (1-R_i) = 2  \qquad {\rm for\,\, each \,\,face}
\eeq
where the sum is over all edges surrounding a given face. These constraints will get
a nice geometric interpretation in section \ref{section_rcharges}.

Let $F$ denote the number of faces, $E$ the number
of edges and $V$ the number of vertices in the brane tiling.
By summing equation (\ref{wcharge}) over the nodes, we get $2 \sum_{edges} R_i=2V$.
Using this and summing equation (\ref{wch20}) over all the faces in the tiling we arrive at the {\bf Euler formula}
for a torus:
\beq
  F - E + V = 0
\eeq
This is a non--trivial statement about the quiver theory which was first observed in \cite{Kol:2003} and derived in \cite{Franco:2005rj}.

In our case the linear 't~Hooft anomaly vanishes \cite{Benvenuti:2004dw}: $Tr R=\sum \mbox{beta functions}=0$, 
so we have to maximize the following function:
\be
  a(R_i) = \frac{9}{32}\sum_i (R_i-1)^3
\label{aexpr}
\ee


The computation of a--maximization for a given quiver gauge theory has by now turned into a standard 
procedure for solving for supersymmetric gauge theories. Furthermore, it serves as good probe for 
consistency checks on quiver theories. Indeed, while there are many theories for which this procedure leads to nice and impressive results, it turns out that there is a large class of quiver gauge theories 
for which a straightforward application of a--maximization gives rise to negative or zero R--charges. This obviously indicates some sign of inconsistency.
Such theories were termed in \cite{Aspinwall:2004vm} as having tachyons, in \cite{Feng:2002kk} as fractional Seiberg duals and in \cite{Herzog:2004qw} as mutations.
All these examples share the same property of having negative R--charges. 

\section{Isoradial embeddings and R-charges}
\label{section_rcharges}

Let us consider again the constraints for the R--charges (section \ref{section_amaxi}):
\beq
\sum_{ i \in edges \,\, around \,\, node } R_i =2 \qquad \qquad {\rm for\,\, each \,\,node}
\eeq
\beq
2 + \sum_{i \in edges\,\, around \,\, face} (R_i-1) =0  \qquad {\rm for\,\, each \,\,face}
\eeq
After multiplying both equations by $\pi$ and rearranging the second one, we arrive at
\be
\sum_{ i \in edges \,\, around \,\, node } (\pi R_i) =2\pi \qquad\qquad {\rm for\,\, each \,\,node}
\label{wch1}
\ee
\be
  \sum_{i \in edges\,\, around \,\, face} (\pi R_i) = (\#edges-2)\pi  
  \qquad {\rm for\,\, each \,\,face}
\label{wch2}
\ee
Now, if we think of $\pi R_i$ as an angle, then we see that the first equation is
just the statement that the angles around a node sum up to $2\pi$, whereas the second equation
tells us that the sum of the internal angles in a polygon is $(\# edges-2)\pi$.

\begin{figure}[ht]
  \epsfxsize = 5cm
  \centerline{\epsfbox{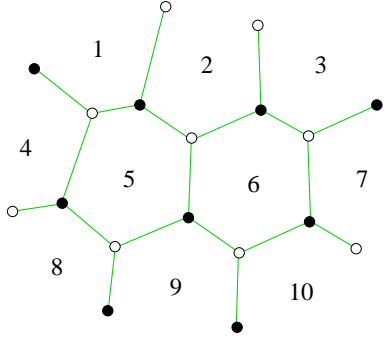}}
  \caption{Isoradially embedded part of an arbitrary brane tiling (in green).}
  \label{isoradi0}
\end{figure}

Where are these angles in the brane tiling? To show this, we need the notion of
isoradial embedding \cite{Duffin1, mercat, Kenyon:2002a}. 
So far the brane tiling was only 
a graph for us, we could freely move around its nodes without causing self--intersection. The 
{\bf isoradial embedding } is an
embedding of the tiling graph into the plane, where the nodes of each face are on a circle
of unit radius. (The edges of the tiling are straight lines).
The square lattice for the conifold provides a trivial example (Figure 4 (i)), where
the unit circles are just the circumcircles of the squares in the tiling.
The squares are of same size so the circumcircles will have the same radius which can be
chosen to be one.

\comment{
\begin{figure}[ht]
  \epsfxsize = 7cm
  \centerline{\epsfbox{isoradi1.eps}}
  \caption{Circumcircles around the faces (in black).}
  \label{isoradi1}
\end{figure}

\begin{figure}[ht]
  \epsfxsize = 7cm
  \centerline{\epsfbox{isoradi2.eps}}
  \caption{The corresponding rhombus lattice (in red).}
  \label{isoradi2}
\end{figure}
}

\begin{figure}[ht]
  \epsfxsize = 15cm
  \centerline{\epsfbox{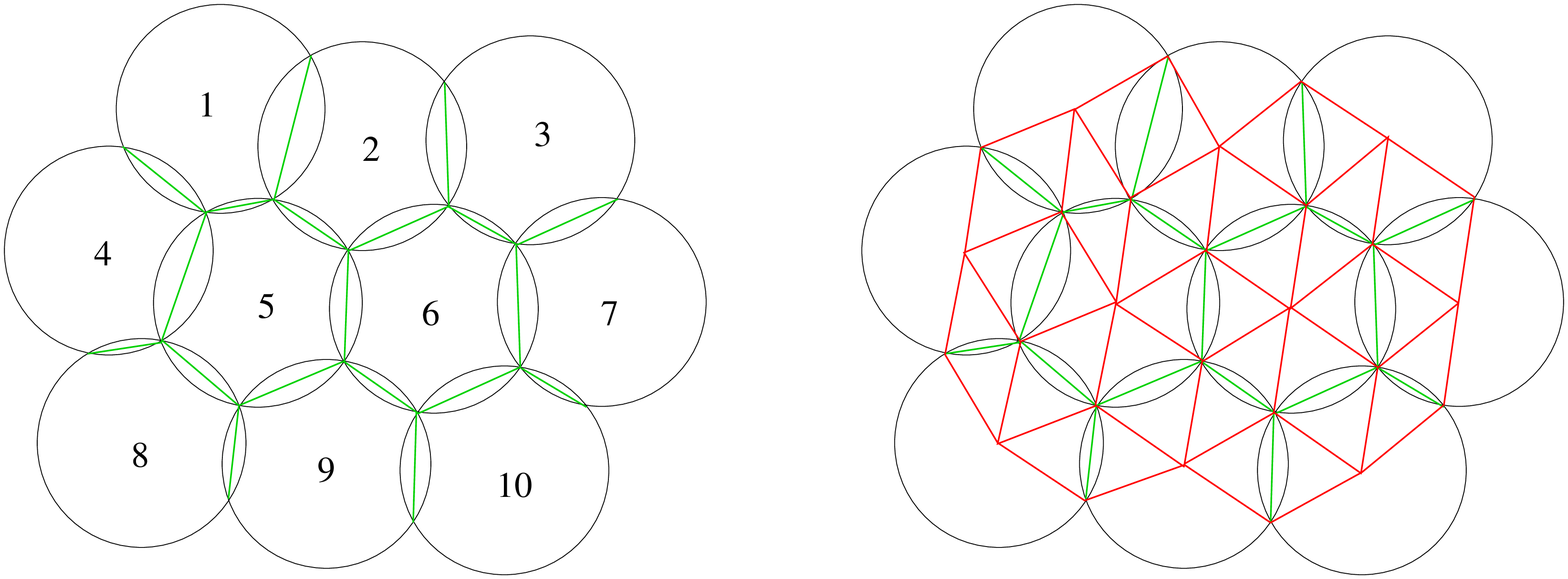}}
  \caption{(i) Circumcircles around the faces (in black), (ii) and the corresponding rhombus lattice (in red).}
  \label{isoradi12}
\end{figure}

To demonstrate a non--trivial example, \fref{isoradi0} shows a small part of a brane tiling\footnote{
From now on, green lines will always denote edges in the brane tilings, 
red lines are edges of the rhombus lattice and (directed) blue lines denote the rhombus loops.}. 
This tiling graph is isoradially embedded into the plane.
This can be seen in \fref{isoradi12} (i), where the black circles are the circumcircles of unit radius 
of the faces in the tiling. The nodes of the brane tiling are sitting at the intersection points
of the circles.

Once we have the tiling isoradially embedded, we can immediately draw the corresponding
{\bf rhombus lattice}\footnote{Also known as quad--graph or diamond lattice.} (\fref{isoradi12} (ii) shows the rhombus lattice in red),
which can be obtained by simply connecting the center of the circles with the nodes
of the face in the brane tiling.
The rhombi (a.k.a. ``diamonds'' in \cite{Kenyon:rhombic}) in this lattice have edges of unit length.
This is guaranteed by the equality of the radii of the circles.
We see that by isoradially embedding our original tiling we gain a lattice of rhombi.
The bifundamental fields of the quiver theory (i.~e. edges
in the brane tiling) are in one--to--one correspondence with the rhombi of this rhombus lattice.

\begin{figure}[ht]
  \epsfxsize = 6cm
  \centerline{\epsfbox{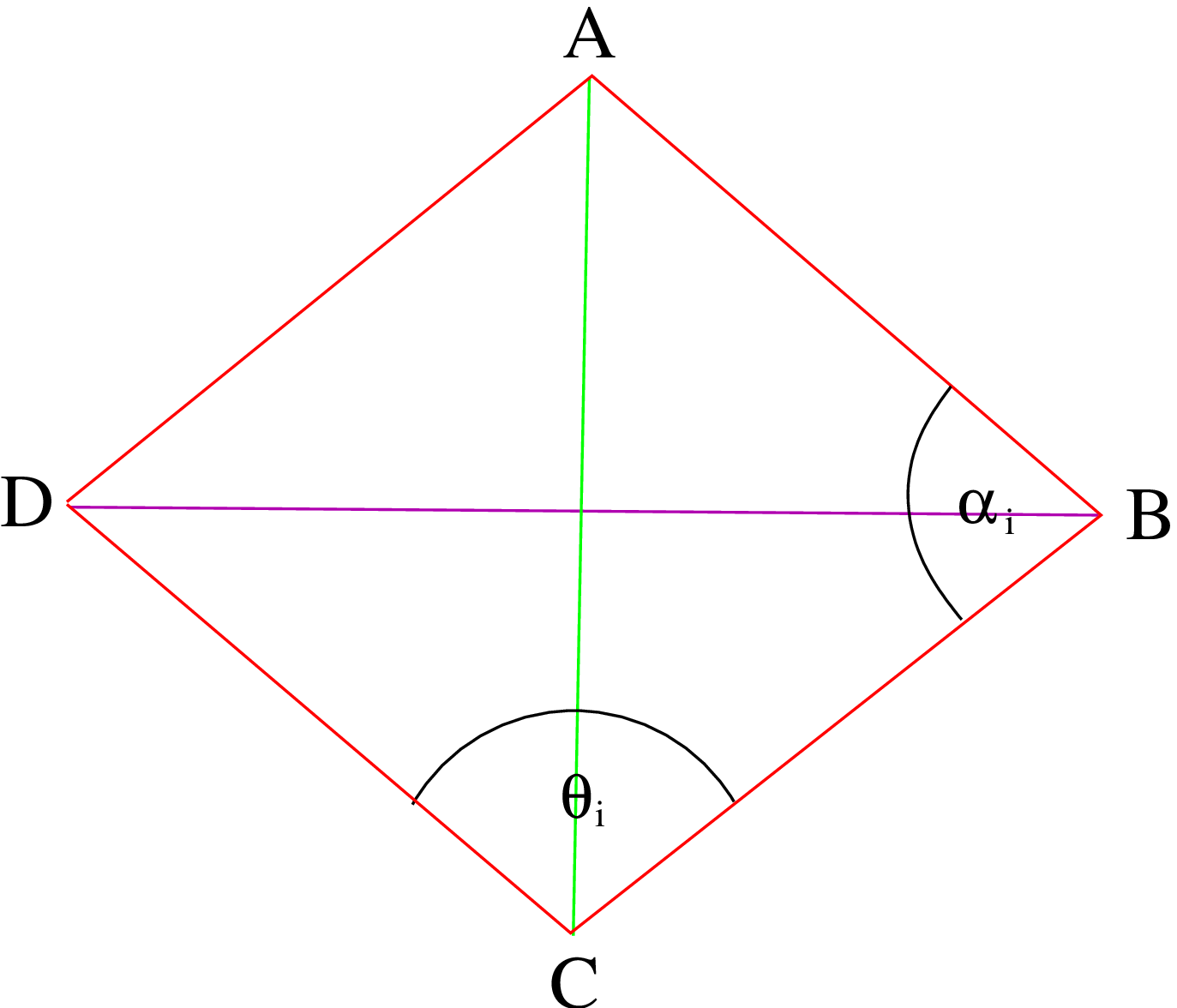}}
  \caption{A rhombus in the lattice. The green line is an edge in the brane tiling,
the magenta line is the corresponding bifundamental field in the periodic quiver.}
  \label{rhombus1}
\end{figure}

Let us study now a single rhombus that is shown in \fref{rhombus1}. 
The green bifundamental edge ($AC$ in \fref{rhombus1}) is just one of the diagonals of the rhombus.
If instead of the green lines we draw the flipped magenta ones 
($BD$ in \fref{rhombus1}) into the rhombus lattice,
then we obtain the dual graph to the tiling, the periodic quiver
(which is also isoradially embedded).
We immediately see that on the level of the rhombus lattice, the quiver and the brane tiling 
are on the same footing. 

In the figure, $\theta_i$ denotes the $DCB$ and $BAD$ angles in the rhombus. 
The shape of the rhombus is characterized by this single angle.
We are now in the position to visualize the R--charges if we set
\be
\theta_i \equiv \pi R_i
\ee
We see that the condition for vanishing beta function to superpotential terms, equation \eqref{wch1} says that the angles around a node in the brane tiling sum up to $2\pi$,
whereas the condition for vanishing beta function to gauge groups, equation \eqref{wch2} is equivalent to the statement that the sum of the internal angles 
of each face in the tiling is $(\# edges-2)\pi$. This is certainly true for a {\bf flat torus}.

It is not {\it a priori} clear that an arbitrary brane tiling graph can be isoradially embedded into the plane.
If the exact R--charges are strictly greater than zero and less than one, then they provide a good
embedding of the rhombus lattice, hence an isoradial tiling.
If some $R_i=0 \ (\mbox{or} \ 1)$, then $\theta_i=0 \ (\mbox{or} \ \pi)$, that is the corresponding rhombus becomes
degenerate. 

The results of this section is that we identified the R--charges of the bifundamental fields with certain angles in the brane tiling.
For any periodic embedding of the rhombus lattice of the brane tiling into the plane
the trial R--charges (defined by the  $\theta_i$ angles in the rhombi) automatically satisfy
the equations \eqref{wch1} and \eqref{wch2}, and vice versa, the set of exact R--charges of the quiver
gauge theory gives a good rhombus lattice and thereby an isoradial embedding of the brane tiling.


Finally, let us transform equation \eqref{aexpr} into the following form using the angles
in \fref{rhombus1}:
\be
  a = -\frac{9}{32\pi^3}\sum_i \alpha_i^3
\ee
Here we used the fact that $\alpha_i = \pi-\theta_i = \pi(1-R_i)$.
The parameter space of the different possible embeddings of the rhombus lattice is nothing,
but the manifold over which one has to do a--maximization. This space will be investigated
in the next section. 

\section{Rhombus loops and zig--zag paths}
\label{section_rloops}

In the last section we introduced a very special type of embedding of the tiling,
the so--called isoradial embedding. This has been used to visualize the R--charges of the
bifundamental fields. In this section we go further and develop new mathematical concepts that will
allow us to study the moduli space of isoradial embeddings that is the parameter space
of a--maximization. 

\begin{figure}[ht]
  \epsfxsize = 11cm
  \centerline{\epsfbox{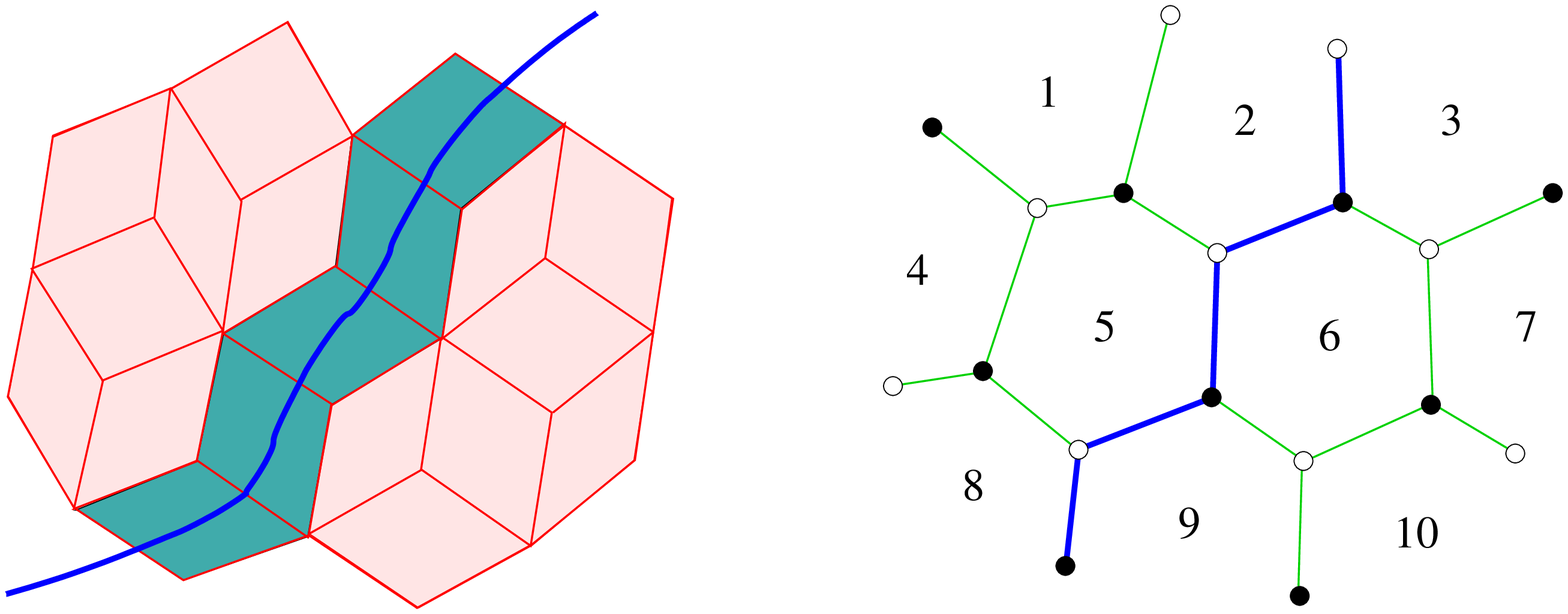}}
  \caption{(i) Rhombus path in the rhombus lattice. 
(ii) Equivalent zig--zag path in the brane tiling. 
We will use blue lines to depict rhombus loops schematically.
The edges which are crossed by the blue line in (i) are all parallel. Their orientation can be
described by an angle, the so--called rhombus loop angle.}
  \label{rloop1}
\end{figure}

The most important new concept that we will continuously use in the present paper is
the notion of the rhombus path (a.k.a. ``train track'' \cite{Kenyon:rhombic}).
A {\bf rhombus path} is defined in the rhombus lattice as a path on rhombi which ``does not turn'', 
i.~e. after entering to a rhombus on one edge, we are exiting on the opposite side
(see \fref{rloop1}). We can assume that the rhombus path is extended to its maximal size, 
which means that in a rhombus lattice on the
surface of $\IT^2$ (or, equivalently, in the periodic rhombus lattice)
it is a closed loop, the {\bf rhombus loop}. The rhombus loops will be of great importance
in the Fast Inverse Algorithm in section~\ref{section_fia}.

\begin{figure}[ht]
  \epsfxsize = 17cm
  \centerline{\epsfbox{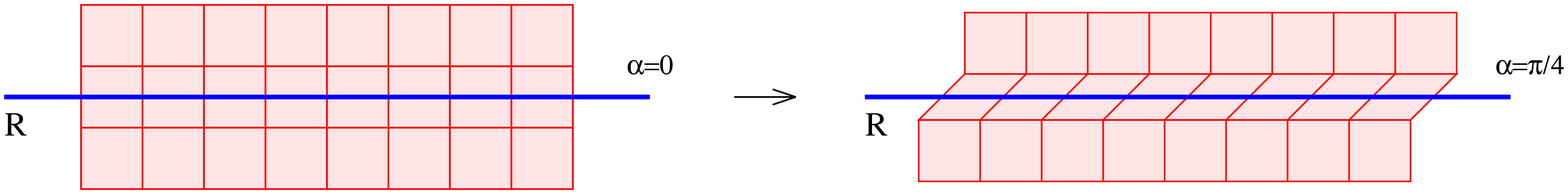}}
  \caption{Tilting along the horizontal $R$ rhombus loop.
The rhombus loop angle $\alpha$ changes during the Dehn--twist.
Here we have chosen $\alpha=0$ to be the vertical direction ($|$), 
hence $\alpha=\pi/4$ corresponds to the  skew edges ($/$).}
  \label{rlooptilting}
\end{figure}

The rhombus edges we are crossing while going along the rhombus loop are all parallel.
Their direction, which can be parametrized by a characteristic angle, 
the {\bf rhombus loop angle} ($\alpha$ and $\beta$ in \fref{rcross}).
This angle can be changed by {\bf tilting} the rhombus loop as in \fref{rlooptilting}.

In \cite{Kenyon:rhombic} it was shown that: (i) No rhombus path
crosses itself (or it is periodic), and (ii) two distinct rhombus paths cross each
other at most once.
These conditions are not always true in our case, because we allow the existence of 
{\bf degenerate rhombi}.
Two--valence nodes also result in collapsing rhombi, they have to be integrated out
before drawing the rhombus lattice.

If the R--charge of a bifundamental field is one, the corresponding rhombus collapses ($\theta_i = \pi$).
This happens for example in the square--octagon phase of the zeroth Hirzebruch surface.
We can also squash the rhombus in the perpendicular direction, if we set the R--charge
equal to zero. As opposed to the $R_i=1$ situation, 
this case is not allowed, it leads to the so--called tachyonic quivers.

If we color the edges in the brane tiling corresponding to the rhombus loop 
(the blue lines in \fref{rloop1}),  we get the so--called {\bf zig--zag path} \cite{Kenyon:2002a}.
This is a path in the tiling which turns maximally left at a node, then maximally
right at the next node, then again left, and so on. 
An example is presented in \fref{rloop1}. The first picture shows the rhombus path,
the second one is the corresponding zig--zag path in the brane tiling.
See \fref{mSPP3} for another example in SPP. Here the blue zig--zag path is periodic.

The zig--zag paths and the rhombus loops are equivalent, the only difference is that they
refer to the same path in different lattices.
Henceforth we will use both terms depending on the context.
At first, it might be non--trivial to understand why there are exactly two zig--zag paths going 
through each tiling edge. This is best seen in the 
rhombus lattice where these two paths are the two ``perpendicular'' rhombus
loops that are crossing the corresponding rhombus.

The zig--zag paths in the tiling are in one--to--one correspondence with zig--zag paths 
in the periodic quiver. These paths in the quiver are oriented loops
hence there are gauge--invariant trace operators that can be constructed by multiplying the
bifundamentals one after the other along the path.
Such an operator is called the {\bf zig--zag operator}.

\subsection{Inconsistent theories}
\label{section_incon}

In the previous sections we reviewed the construction of brane tilings, visualized R--charges
as certain angles in the tiling and introduced the new concept of zig--zag paths.
One may now imagine that for any arbitrary bipartite tiling there exists a corresponding quiver theory. 
Unfortunately, this is not the case and there exist some bipartite graphs which do not give 
meaningful quiver theories.
So far in the literature there was no other restriction on consistent tilings, than bipartiteness. 
In this paper we are going to give a simple constraint that has to be satisfied by every
consistent brane tiling.

One can construct the $Y_6$ Calabi--Yau manifold as a \kahler quotient \cite{Kronheimer:1989zs} 
that is as the (classical) vacuum moduli space of a gauged linear sigma model (GLSM)
\cite{Witten:1993yc,Douglas:1997de}.
The ``Fast Forward Algorithm''  (\cite{Franco:2005rj}, see also \cite{Hanany:2005ve})
computes the toric diagram of the singular Calabi--Yau from the brane tiling. 
The algorithm also gives the multiplicities of the GLSM fields, these appear in the toric diagram.
It is possible that from a given tiling the Fast Forward Algorithm produces a toric diagram,
whose area is smaller than what we expect from the number of the
corresponding gauge theory. This is a good sign of inconsistency of the theory.
Then, typically, a--maximization gives zero R--charges for some of the bifundamental 
fields\footnote{In \cite{Franco:2004wp} such tachyonic quivers were investigated in the
context of $(p,q)$--webs.}.
For such theories, we also get GLSM field multiplicities in the corners of the toric diagram
(see \fref{hirze}). These {\bf external multiplicities} should be further investigated.

\begin{figure}[ht]
  \epsfxsize = 5cm
  \centerline{\qquad \qquad\epsfbox{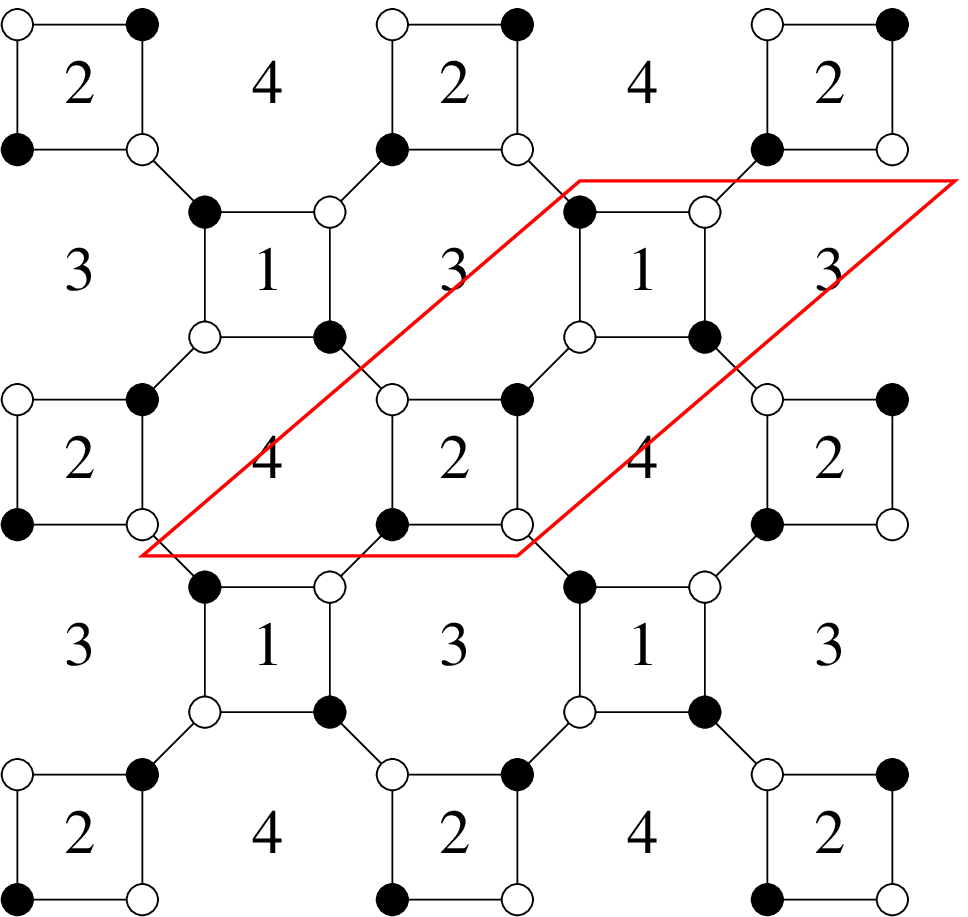}}
  \caption{Hirzebruch zero brane tiling.}
  \label{f0t}
\end{figure}

Partial resolutions of the singularity correspond to turning on Fayet--Iliopoulos terms 
in the supersymmetric gauge theory side
and leads to {\bf Higgsing} in the quiver gauge theory. 
The FI terms govern the size of the blow--ups.
The effective theory at scales smaller than the expectation value of the Higgsed field
can be described by the Higgsed quiver and superpotential \cite{Feng:2000mi,Feng:2002fv}.
Here we consider the inverse of this process, the so--called {\bf un--Higgsing}.
In the level of brane tiling this can be implemented by adding a new edge to the graph.
This edge divides a face into two faces, therefore the number of gauge groups
increases by one, the number of bi-fundamental fields increases by one while the number of terms in the superpotential remain the same. Alas, not all possible un--Higgsings of the theory are consistent,
in fact, it is a non--trivial problem to determine the allowed un--Higgsings 
for a given brane tiling.

To demonstrate consistent and inconsistent un--Higgsing, we consider the Hirzebruch zero ($F0$) surface.
$F0$ has two toric phases that are connected by Seiberg duality. The brane tiling for one of the phases is 
the square lattice. We will study the other phase that is
the square--octagon lattice which is depicted in \fref{f0t}. 
We consider two possible un--Higgsings of the theory that
are shown in \fref{hirzetilings}. The new edge (dashed line) is dividing the original
face $4$ into two faces $4$~\&~$5$.
The first un--Higgsing (i) leads to an inconsistent theory. 
By means of the Fast Forward Algorithm we can compute its toric diagram
with the multiplicities of the GLSM fields. The results are shown in \fref{hirze}.
We see that during the un--Higgsing the area of the diagram remained the same, meanwhile an external
multiplicity (the $3$ in the corner) appeared.

\begin{figure}[ht]
  \epsfxsize = 10cm
  \centerline{\qquad \qquad\epsfbox{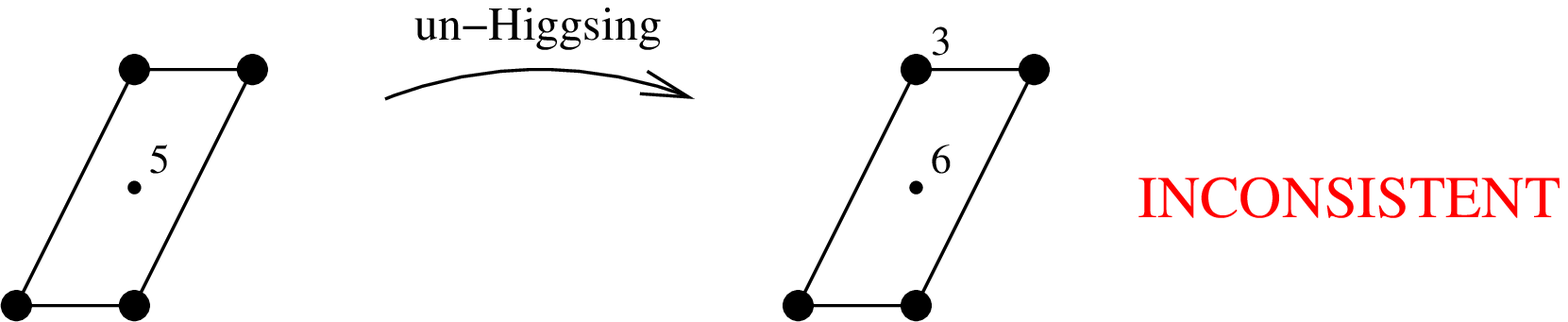}}
  \caption{(i) Hirzebruch zero toric diagram  (ii) un--Higgsed Hirzebruch. The area remains the same,
external multiplicities appear.}
  \label{hirze}
\end{figure}

\begin{figure}[ht]
  \epsfxsize = 10cm
  \centerline{\qquad \qquad\epsfbox{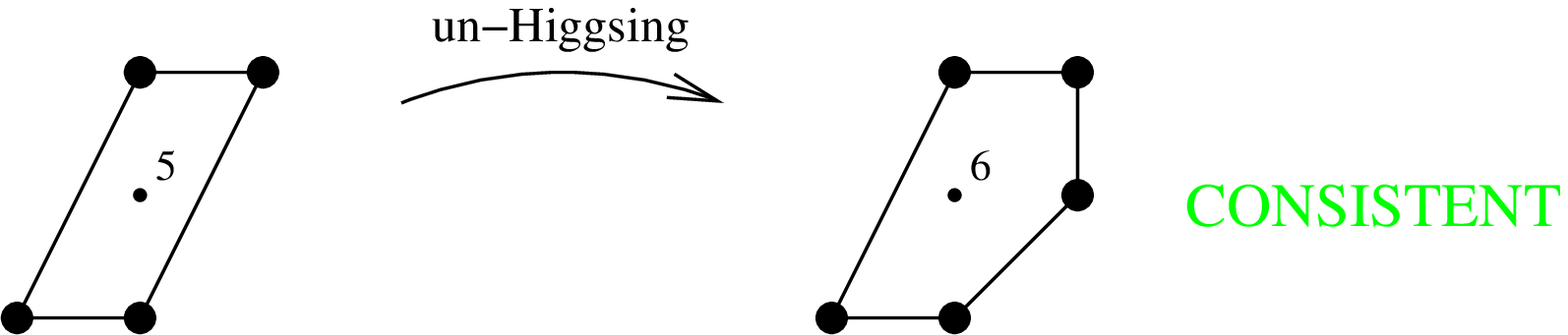}}
  \caption{(i) Hirzebruch zero toric diagram  (ii) un--Higgsed Hirzebruch. The area 
increases by $1/2$ corresponding to the new face in the brane tiling.}
  \label{hirze2}
\end{figure}

We now consider another un--Higgsing that adds the line with a different orientation (see \fref{hirzetilings} (ii)).
This theory is consistent. The corresponding toric diagram is shown in \fref{hirze2}.


\begin{figure}[ht]
  \epsfxsize = 13cm
  \centerline{\epsfbox{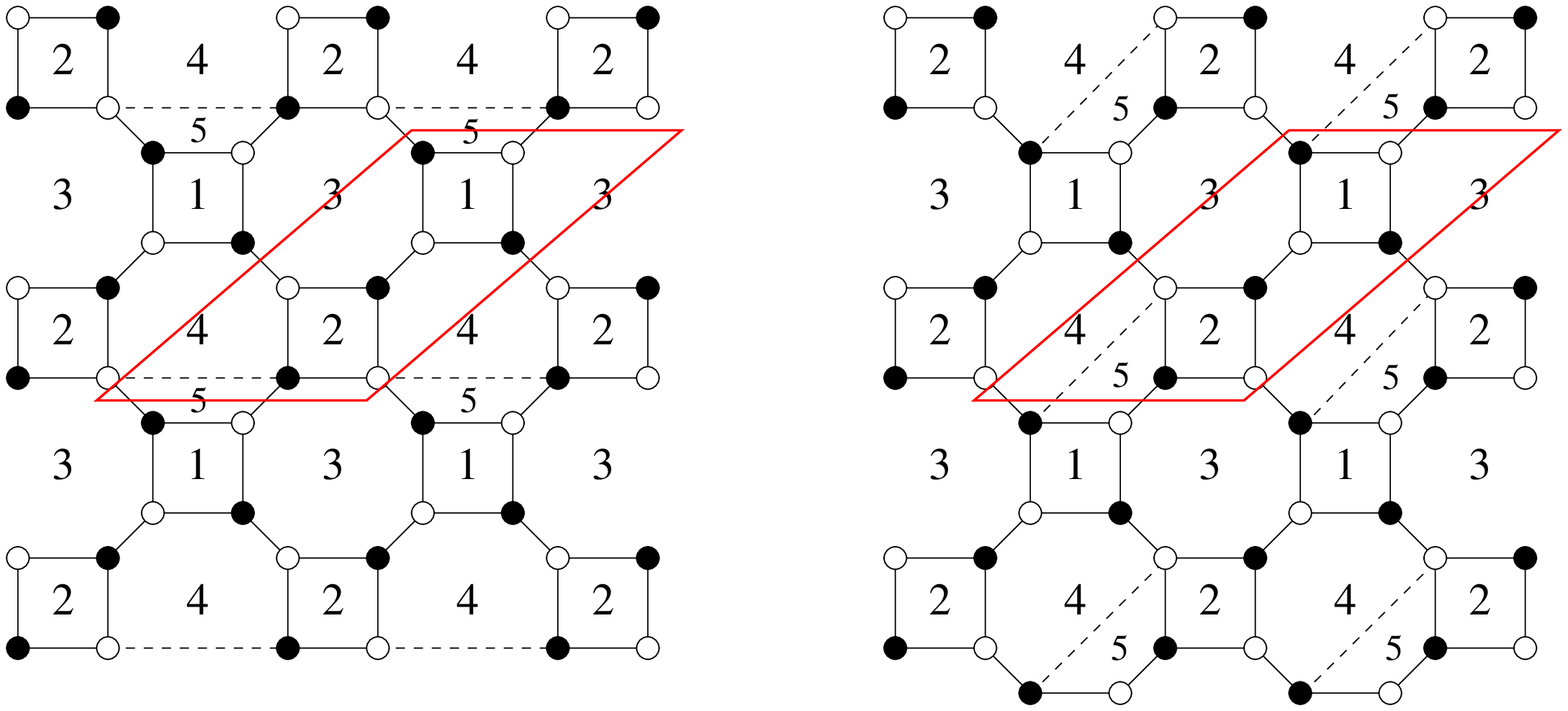}}
  \caption{(i) Hirzebruch zero inconsistently un--Higgsed. (ii) Consistent un--Higgsing.}
  \label{hirzetilings}
\end{figure}


\begin{figure}[ht]
  \epsfxsize = 14cm
  \centerline{\epsfbox{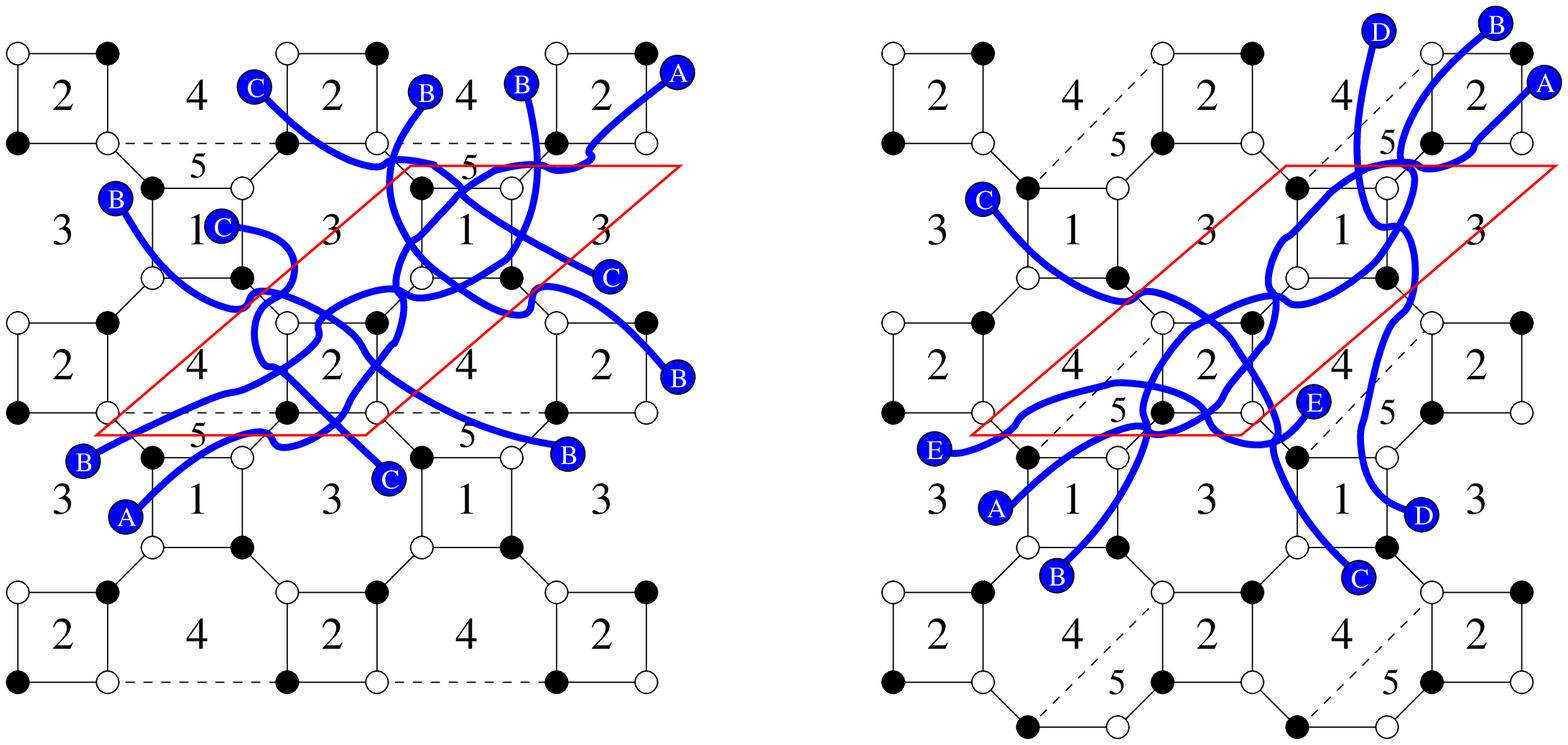}}
  \caption{(i) Inconsistently un--Higgsed Hirzebruch. 
The rhombus loops are indicated with the blue lines. 
The zig--zag paths contain the edges that are crossed by the blue paths.
The following rhombus loops 
are obtained: $A:(0,-1)$ \ $B:(-2,2)$ \ $C:(2,-1)$.
Here $(a,b)$ denotes the homology class of the path.\newline
(ii) Consistently un--Higgsed F0. The rhombus loops reproduce the $(p,q)$--legs of the toric
diagram (Figures 6, 7): $A:(0,-1)$ \ $B:(0,1)$ \ $C:(-2,1)$ \ $D:(1,-1)$ \ $E:(1,0)$.}
  \label{F0RL}
\end{figure}

In the Fast Forward Algorithm it is {\it a priori} unclear why these small changes in the tiling 
lead at one time to a consistent and at another time to an inconsistent theory.
Having discussed the main mathematical concepts that we need,
we can now understand what causes the inconsistency.

\fref{F0RL} shows the rhombus loops\footnote{We can choose the {\bf direction} of the rhombus loops so that they pass the
black nodes on the left-hand side.} for the two different un--Higgsing of $F0$.
The blue lines are crossing edges which are the edges of the corresponding zig--zag paths.
The pictures show the rhombus loops only inside the fundamental cell.
For the inconsistent tiling (i) we obtain only three rhombus loops, it does not 
reproduce the $(p,q)$--legs of the toric diagram which we obtained by the Fast
Forward Algorithm (\fref{hirze}). On the other hand, the zig--zag paths of the consistent tiling (ii) 
give the legs properly (\fref{hirze2}).

In the first tiling the edge between face $4$ and $5$ is
at the intersection point of the $B$ loop with itself. The corresponding rhombus
in such cases is always degenerate, 
because all the four edges of the rhombus must be parallel,
therefore the first tiling is inconsistent.
We can state this in general: {\bf Self--intersecting zig--zag paths lead to inconsistent brane tilings}.

\begin{figure}[ht]
  \epsfxsize = 4cm
  \centerline{\epsfbox{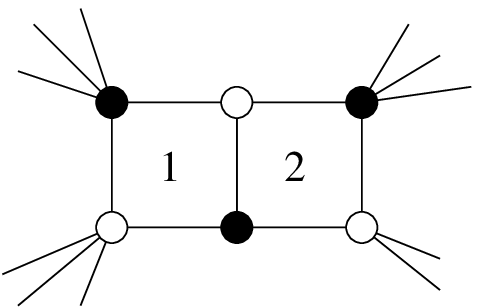}}
  \caption{The subgraph connects to the rest of the tiling through its four nodes in the corner.
  No consistent brane tiling can contain this subgraph, because it results in collapsing rhombi
  and vanishing R--charges.}
  \label{incons}
\end{figure}

This example demonstrated how zig--zag paths can be used to determine whether the tiling
is consistent or not. Besides these computations,
the rhombus loop technique enables us to generate simple rules that must be
satisfied by any consistent tiling. This might help in the construction of such tilings.
An example is presented in \fref{incons}.
This subtiling cannot be part of any consistent tiling.
It is clear from the corresponding
(degenerate) rhombus lattice that there are zero R--charges
as the reader may check. The inconsistency can be also seen
by performing Seiberg duality on face~$1$ that creates a face with only two edges.

\subsection{Conjecture of $(p,q)$--legs and rhombus loops}
\label{conj}

In the previous section we investigated rhombus loops in the rhombus lattice and equivalent zig--zag paths
in the brane tiling. We have seen that one can use these paths to decide whether the tiling is {\it a priori}
consistent or not (i.~e. before doing a--maximization).
In the followings we make an observation which will enable us to develop the Fast Inverse Algorithm
in section \ref{section_fia}.

\begin{figure}[ht]
  \epsfxsize = 9cm
  \centerline{\epsfbox{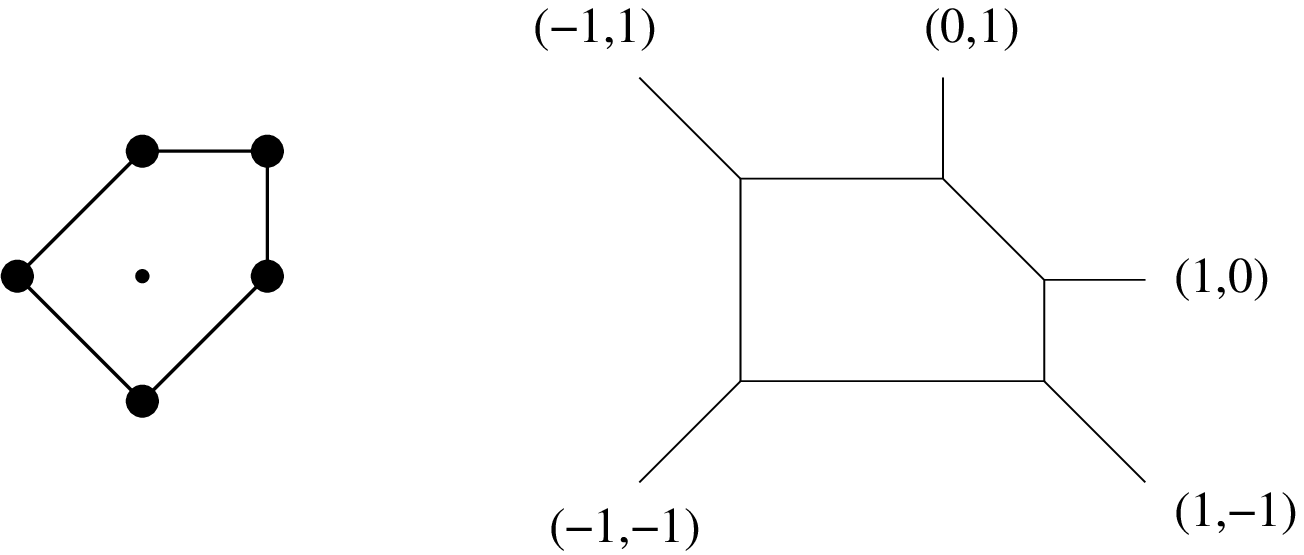}}
  \caption{Toric diagram (i) and $(p,q)$--web (ii) for del Pezzo 2.
The charges of the external branes are shown. According to the conjecture, these correspond to the 
homology classes of the rhombus loops
in the brane tiling.}
  \label{spq}
\end{figure}

To state the conjecture we introduce the notion of $(p,q)$--webs.
{\bf $(p,q)$--webs} were introduced in \cite{Aharony:1997ju} to study five dimensional
gauge theories with 8 supercharges (i.~e. $\mathcal{N}=1$). The $(p,q)$--web describes a configuration of 5--branes
in Type IIB string theory. These webs might be interpreted as
``dual graphs'' to toric diagrams as it was noticed in \cite{Aharony:1997bh}.
This observation has been proven in \cite{Leung:1997tw}.
An example is shown in \fref{spq}.
The geometry can be described by a $\IT^2$ fibration
over the web. A circle in the fibre degenerates at each line of the diagram
and at the nodes the whole fibre collapses. The lines of the web have rational slopes
denoted by two integers: $(p_i,q_i)$. These are the $(p,q)$ charges of the branes.
A D5--brane is assigned a $(1,0)$ charge whereas the NS5--brane carries $(0,1)$ charge.
These two type of branes correspond to horizontal and vertical lines in the web.
At each node we have three branes intersecting each other and their charges must sum up to zero:

\be
\sum_{i} p_i = 0 \qquad \sum_{i} q_i = 0 
\ee


In the followings we will use {\bf $(p,q)$--legs}. These are the external lines in the
$(p,q)$--web and they extend to infinity. 
Their direction is perpendicular to the corresponding edge of the (dual) toric diagram.

An important observation is that {\bf for each rhombus loop of homology class $(p,q)$
there is a corresponding $(p,q)$--leg in the toric diagram}. We will heavily use this 
in section \ref{section_fia}. The conjecture has been checked for many consistent
brane tilings. Inconsistent tilings tipically do not satisfy this criterion.
By reading off the zig--zag paths from the tiling we might arrive at the toric data
faster than by the usual Kasteleyn matrix process \cite{Franco:2005rj, Hanany:2005ve}.
We simply need to draw all the zig--zag paths (each edge has two of them)
and from their homology classes the $(p,q)$--legs are obtained.
These legs uniquely determine the toric diagram of the Calabi--Yau cone.

\begin{figure}[ht]
  \epsfxsize = 3cm
  \centerline{\epsfbox{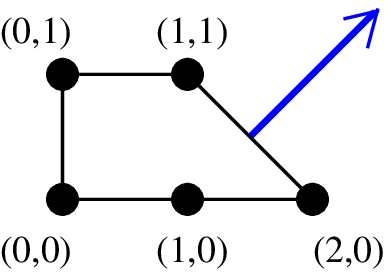}}
  \caption{Toric diagram for the SPP. We have drawn the blue $(p,q)$--leg 
between the nodes  $(1,1)$ and $(2,0)$.
The zig-zag path corresponding to the leg
is shown in Figure~19.}
  \label{mSPP2}
\end{figure}
\begin{figure}[ht]
  \epsfxsize = 15cm
  \centerline{\epsfbox{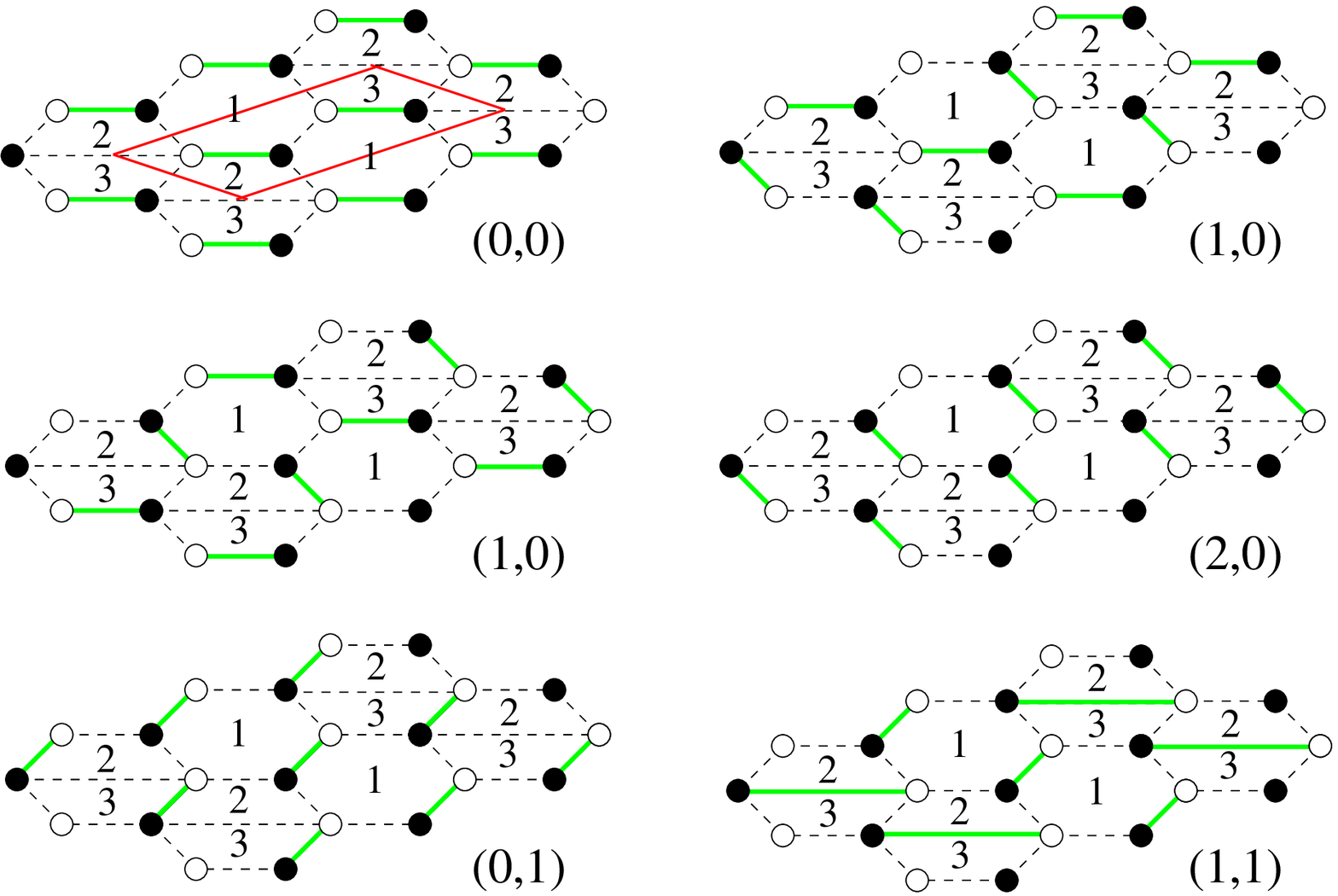}}
  \caption{The six periodic perfect matchings of SPP \cite{Franco:2005rj}. 
The green edges are contained in the matching, the dashed lines are the other edges
of the tiling.
The $(s,t)$ numbers are the corresponding points in the toric diagram (Figure~17).}
  \label{mSPP}
\end{figure}

Another observation\footnote{The results of the rest of the paper will not depend on this observation.} 
is that we can generate zig--zag paths by means of perfect matchings.
A {\bf perfect matching} is a subgraph of the tiling which contains all the nodes and
each node has valence one \cite{Kenyon:2003uj, Kenyon:2002a}. This means that a perfect
matching is a set of {\bf dimers} (edges in the brane tiling) that are separated,
i.~e. they don't touch each other, furthermore they cover all the nodes. Therefore, 
we have altogether $V/2$ dimers in each perfect matching, where $V$ denotes the number
of nodes in the tiling.
To demonstrate this, we have drawn the periodic perfect matchings for the Suspended Pinch Point
(\fref{mSPP}) whose toric diagram is shown in \fref{mSPP2}.

It can be easily checked by the reader that if we put two perfect matchings 
$A$ and $B$ on top of each other (this is denoted by $A+B$), 
then we obtain loops and separate edges which we neglect.
Let us fix a reference perfect matching $R$. Now for each matching $A_i$ we can define an integer 
height function. The loops of $R+A_i$ denote the change in the height as in an ordinary map.
The height function is a well--defined function on the infinite periodic tiling, but on the
2--torus it has monodromy that is described by two integers: $(s,t)$.
These numbers are the change in the height as we go along the two non--trivial cycles of the torus
of the brane tiling.
Such pairs are assigned to every perfect matching. For SPP these vectors are shown in \fref{mSPP}.
(Here we used the first perfect matching as a reference matching.)
These pairs are coordinates of points in the toric diagram,
in fact, the toric diagram is the (convex) set of all such points.
The change in the reference matching merely translates the toric diagram.

\begin{figure}[ht]
  \epsfxsize = 7cm
  \centerline{\epsfbox{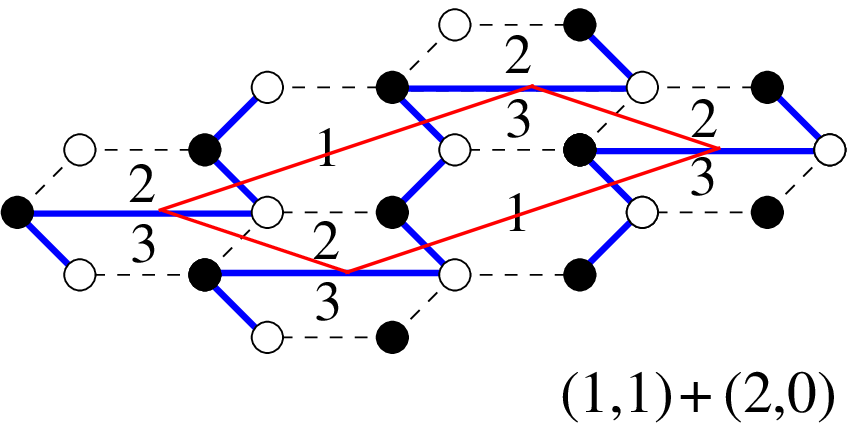}}
  \caption{The $(1,1)$ and $(2,0)$ perfect matchings on top of each other.
We see the emerging  $(1,1)$ homology zig--zag loop which corresponds to the blue $(p,q)$--leg
in Figure~17.}
  \label{mSPP3}
\end{figure}

Now if we choose two adjacent points in the toric diagram then there are perfect matchings
corresponding to them whose superposition is
(experimentally) a zig--zag path. We demonstrate an example for SPP.
The two neighboring matchings have  $(1,1)$ and $(2,0)$ coordinates in the toric diagram.
Their superposition is shown in \fref{mSPP3}. The emerging non--trivial blue cycle
(zig--zag path) has homology $(1,1)$ which precisely corresponds to the blue $(1,1)$ leg
in \fref{mSPP2} which is sitting between the two adjacent points.

For further informations on perfect matchings and the dimer model the reader should refer 
to \cite{Franco:2005rj, Hanany:2005ve, Kenyon:2003uj, Kenyon:2002a}.

\begin{figure}[ht]
  \epsfxsize = 7cm
  \centerline{\epsfbox{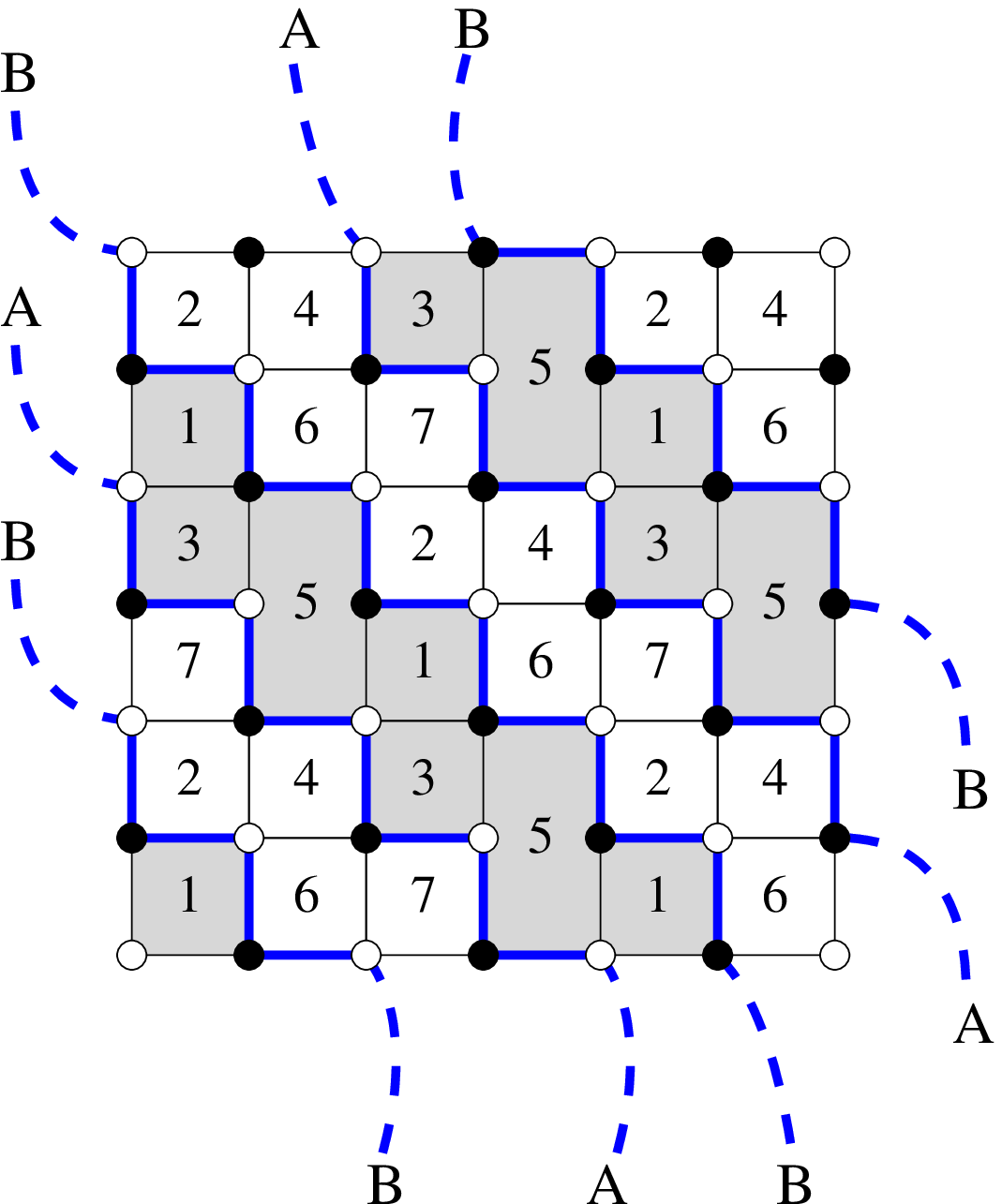}}
  \caption{$PdP_4$ model I brane tiling with a (1,0,1,0,1,0,0) $\mathcal{N}=2$ fractional brane
\cite{Franco:2005zu}.
The bounding rhombus loops ($A$~and~$B$ zig--zag paths) are shown in blue.}
  \label{frac2}
\end{figure}

In a recent paper \cite{Franco:2005zu}, fractional branes were studied in the context
of brane tilings. 
The fractional brane is a D5--brane wrapped on a 2--cycle that vanishes at the tip of the cone.
Adding $M$ fractional branes changes the rank of the $SU(N)$ gauge groups of the quiver.
For deformation branes some of the ranks increase by $M$.
One can shade these tiles as shown in \fref{frac2}.
Zig--zag paths naturally show up as boundaries of these shaded areas.

\subsection{Parameter space of a--maximization}
\label{amaxiparam}

\begin{figure}[ht]
  \epsfxsize = 7cm
  \centerline{\epsfbox{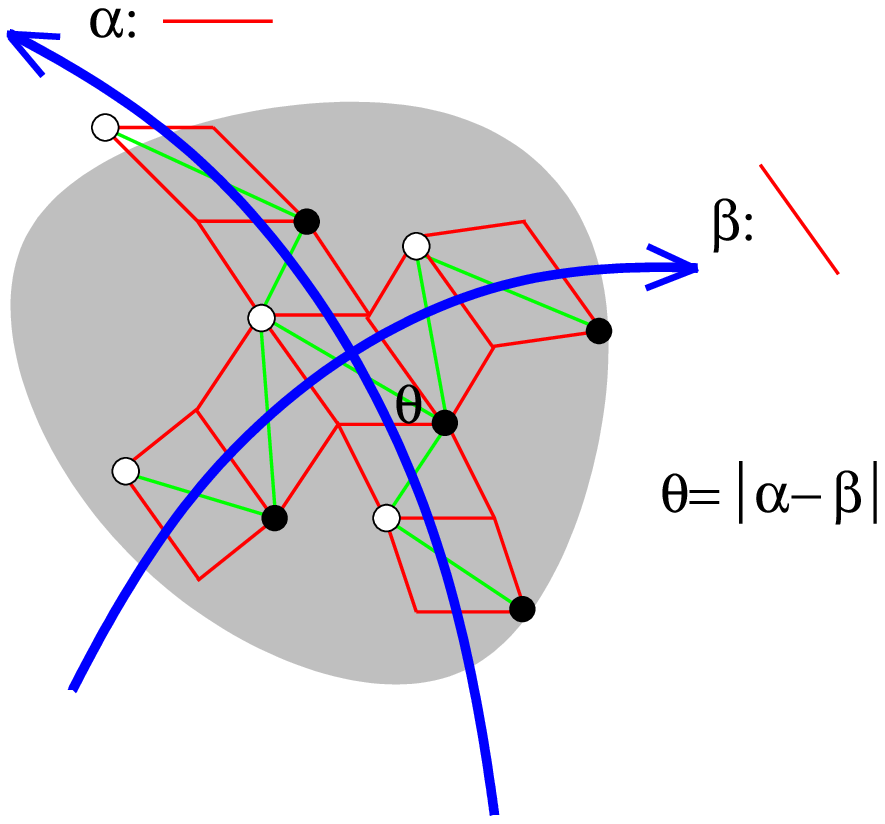}}
  \caption{Assigning angles ($\theta$) to the rhombus loops. 
The figure shows two intersecting blue rhombus paths. There is a single rhombus and a green bifundamental edge 
at the intersection of these paths. This bifundamental has an R--charge that is proportional
to the angle $\theta$ of the rhombus.
This angle is just the difference of the rhombus loop angles $\alpha$ and $\beta$ assigned to the two rhombus paths:
$R\pi = \theta = |\alpha-\beta |$ (or $\pi-|\alpha-\beta |$ depending on the orientation).}
  \label{rcross}
\end{figure}

We have defined the rhombus loop angle that is assigned to a rhombus path. 
This angle gives the relative orientation of the parallel edges in the path.
We have seen that we can tilt the rhombi in a rhombus path by changing its rhombus loop angle (\fref{rlooptilting}).
In fact, we can parametrize the entire space of different embeddings of the rhombus lattice
(i.~e. the isoradial embeddings of the brane tiling) by these rhombus loop angles
\cite{Kenyon:rhombic}.
At the intersection point of two rhombus paths, we find a single rhombus, whose angles ($\theta$ and $\pi-\theta$)
are determined by the difference of the rhombus loop angles of the paths (\fref{rcross}),
because they fix the orientation of the edges of the rhombus.
This angle $\theta$ is proportional to the R--charge of the field sitting in the rhombus
as we have seen in section \ref{section_rcharges}.

This means that we can parametrize the convex polyhedron space \cite{Kenyon:rhombic}
of trial R--charges by the set of rhombus loop angles.
The number of such loops is $d$, which is equal to the number of the edges of the toric
diagram according to our conjecture in section \ref{conj}.
One of the rhombus loop angles can be set to zero by a global rotation of the rhombus lattice.
This reduces the dimension of the parameter space to $d-1$. 
In \fref{rcross} this has already been done, because the $\alpha$ angle is zero 
(the parallel edges in the corresponding rhombus path are horizontal).

Let us see how can we identify the $d-1$ different parameters in the quiver gauge theory:
In the superconformal quiver gauge theory the R--symmetry can mix with every anomaly--free global $U(1)$
symmetry that commutes with charge conjugation.

The global {\bf baryonic $U(1)$}'s are gauge symmetries in the gravity dual picture.
$H_3 (X_5,\Z)=\Z^{d-3}$ (see \cite{newVegh}), i.~e. the number of independent 3--cycles in 
the $X_5$ Sasaki--Einstein
manifold is $d-3$, hence the Kaluza--Klein reduction of the Ramond--Ramond 4--form gives $d-3$ 
different gauge fields in $AdS_5$. These local symmetries correspond in the dual quiver theory 
to global baryonic $U(1)$'s.

Tilting the lattice along a rhombus loop means that the R--symmetry is mixing with a certain $U(1)$ charge. 
The bifundamentals along the loop have $+1$ and $-1$ charges alternatingly under this $U(1)$ and
all the other fields have zero charges.
The baryonic $U(1)$'s are linear combinations of these charges.

We identified $d-3$ degrees of freedom as the mixing of the R--charge with the baryonic charges.
The two remaining charges correspond to the mixing with the two {\bf flavor $U(1)$ charges}. 
These are dual to the Abelian part of the isometry
group of the Sasaki--Einstein manifold which is mixing with the Reeb vector
in the sense of Z--minimization (see \cite{Martelli:2005tp} for details).
The corresponding tiltings are roughly speaking Dehn--twists along the two
nontrivial $(1,0)$ and $(0,1)$ cycles.

One can compute the number of possibly different R--charges for the quiver theory
in the following way. Let us fix two rhombus loops (zig--zag paths).
It is clear that whenever they cross one another, they
produce a bifundamental field with the same R--charge. This follows from the
fact that the rhombus loop angles of the two loops fully determine the orientation of the
rhombus edges, i.~e. they fix the R--charge of the field.
Therefore we can get different charges
only from different rhombus loop intersection points. 
We can count the number of different possible R--charges.
Out of the $d$ loops we are choosing two in all possible ways:
\be
\Big{(} \small{\begin{array}{c} d \\ 2 \end{array}} \Big{)}=\frac{d(d-1)}{2}
\ee
which gives the maximum possible number of different R--charges of the quiver theory.

\section{Fast Inverse Algorithm}
\label{section_fia}


The above discussed techniques based on the isoradial embeddings, rhombus loops 
and zig--zag paths allow us to develop the {\bf Fast Inverse Algorithm},
which constructs the brane tiling from arbitrary toric diagrams.
The brane tiling encodes the quiver (dual graph), the superpotential data (nodes),
hence uniquely describes the quiver gauge theory. 
Therefore, by means of the Fast Inverse Algorithm  {\bf we are able to
compute an AdS/CFT dual to any toric singularity}.
(The algorithm is somewhat complicated by the fact that the resulting theory is highly non--unique.
This phenomenon will be investigated in section \ref{section_dualities}.)

In the following, we describe the algorithm by presenting examples.

\subsection{ $\C^3$ \ ($\mathcal{N}=4$)}

\begin{figure}[ht]
  \epsfxsize = 2.5cm
  \centerline{\epsfbox{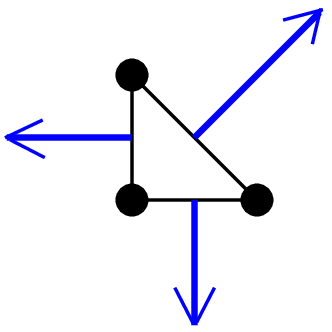}}
  \caption{$\C^3$ toric diagram}
  \label{tdC3}
\end{figure}

In \fref{tdC3} the toric diagram of the flat $\C^3$ Calabi--Yau manifold is shown.
The polygon has three edges, hence three $(p,q)$--legs (the blue arrows) with 
homology classes $(-1,0)$, $(0,-1)$ and $(1,1)$. These correspond to the three rhombus loops
in the rhombus lattice, or to the zig--zag paths in the tiling.

\begin{figure}[ht]
  \epsfxsize = 6cm
  \centerline{\epsfbox{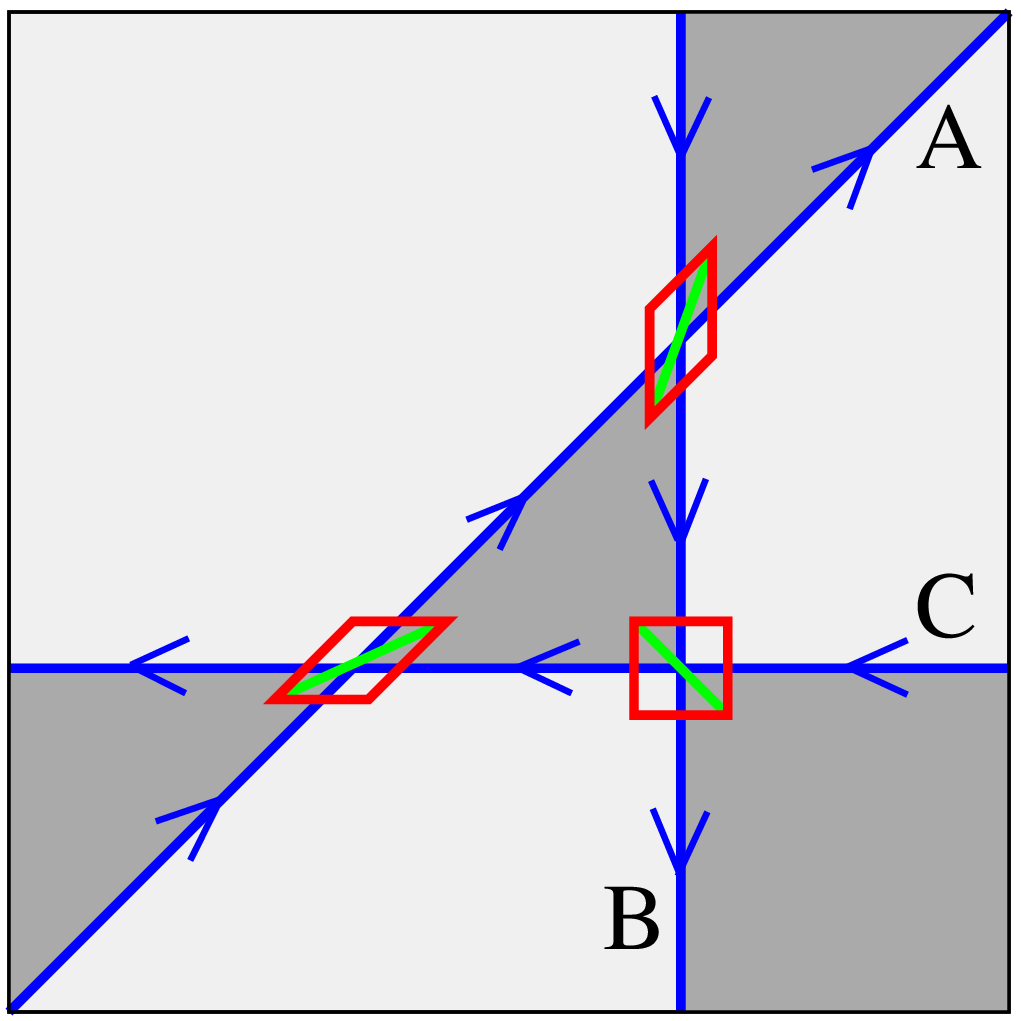}}
  \caption{Rhombus loop diagram of $\C^3$. The blue rhombus loops are the D6--branes. At the intersection
points we get massless fields. The dark faces are terms in the superpotential, the light
faces are the gauge groups. These correspond respectively to nodes and faces in the brane tiling.
The rhombi are shown in red, the brane tiling edges are green.}
  \label{fiaC31}
\end{figure}

\begin{figure}[ht]
  \epsfxsize = 14cm
  \centerline{\epsfbox{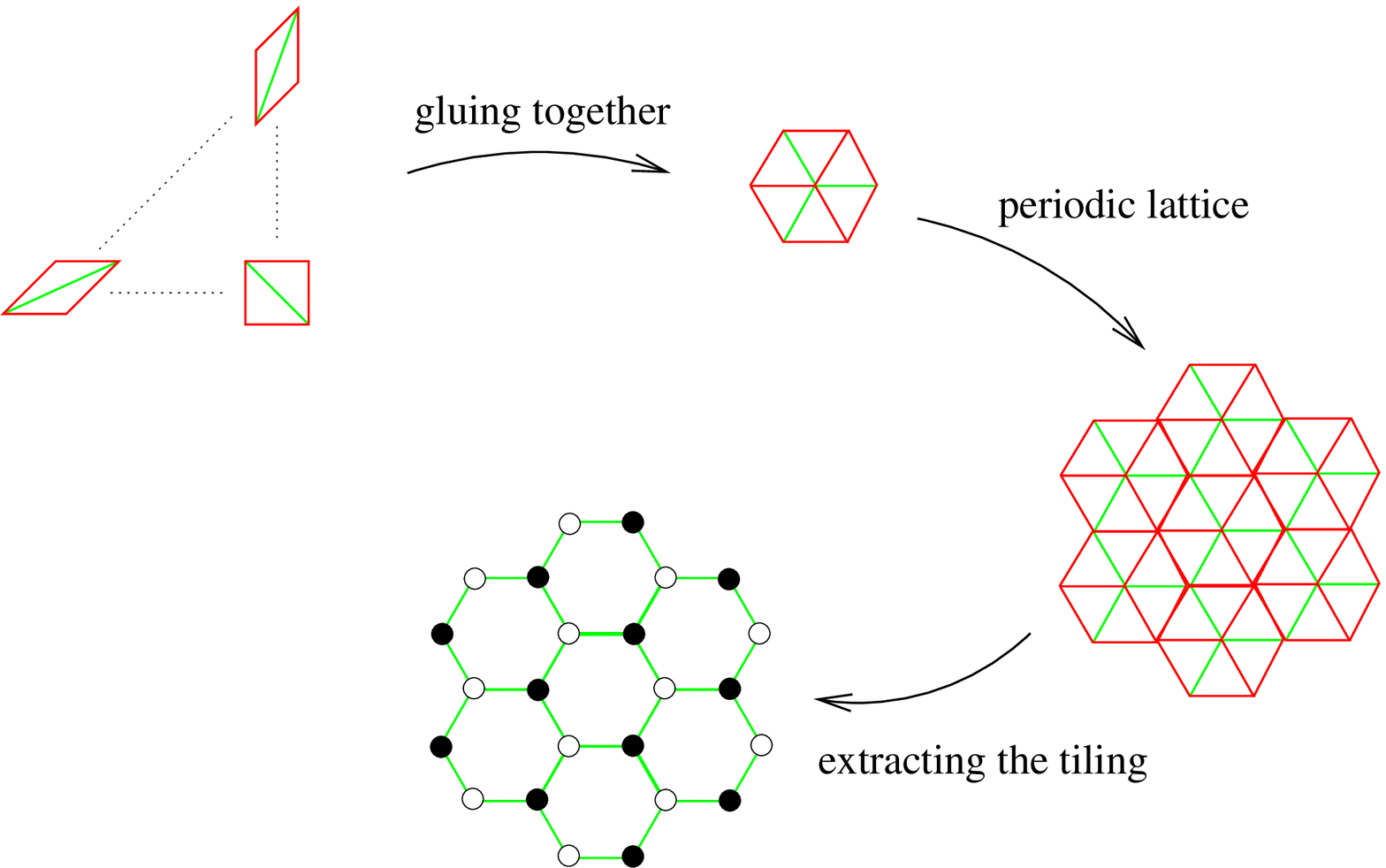}}
  \caption{From the rhombi to the brane tiling. We glue the rhombi together that arise
at the intersections of rhombus loops (Figure 27). We glue the edges that are connected
by the rhombus loops. Each rhombus has a green tiling edge in it, from which we obtain
the entire (hexagonal) brane tiling.
}
  \label{fiaC32}
\end{figure}

We now draw these $(p,q)$ cycles in the fundamental cell (see the blue lines in \fref{fiaC31})
this is the {\bf rhombus loop diagram}.
In the language of the rhombus lattice, at each intersection
point we have a single rhombus, which is shown in red. Each rhombus comes with a single
bifundamental edge in the brane tiling, these edges are shown in green.
To obtain the rhombus lattice, we have to glue these rhombi together (in a periodic fashion),
so that along the blue lines we get rhombus paths (\fref{fiaC32}). 
Once we have the (red) rhombus lattice it is trivial to obtain
the (green) brane tiling which encodes the quiver gauge theory.

We shaded some of the faces in the rhombus loop diagram. From the algorithm it is clear that
these correspond to the (black or white) nodes, whereas the light faces
correspond to the faces in the brane tiling (see also \fref{fiaL131} where the green brane tiling
is drawn directly on top of the rhombus loop diagram of $L^{131}$).
We see that the rhombus loop diagram treats the gauge groups and the terms in the superpotential
on equal footing.

For this simple example, we have only three rhombi, i.~e. three fields, which turn out to
be adjoints, because we have only one gauge group in the tiling.
In \fref{fiaC32} we recover the hexagonal lattice of $\mathcal{N}=4$.

\newpage
\subsection{Conifold}

\begin{figure}[ht]
  \epsfxsize = 2.5cm
  \centerline{\epsfbox{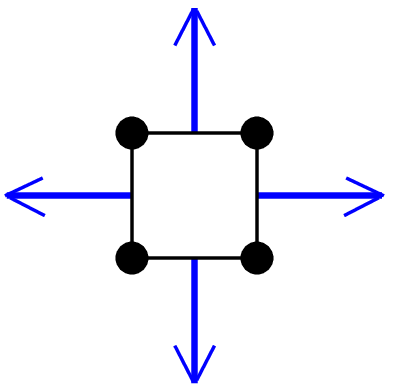}}
  \caption{Conifold toric diagram}
  \label{tdconifold}
\end{figure}

\begin{figure}[ht]
  \epsfxsize = 6cm
  \centerline{\epsfbox{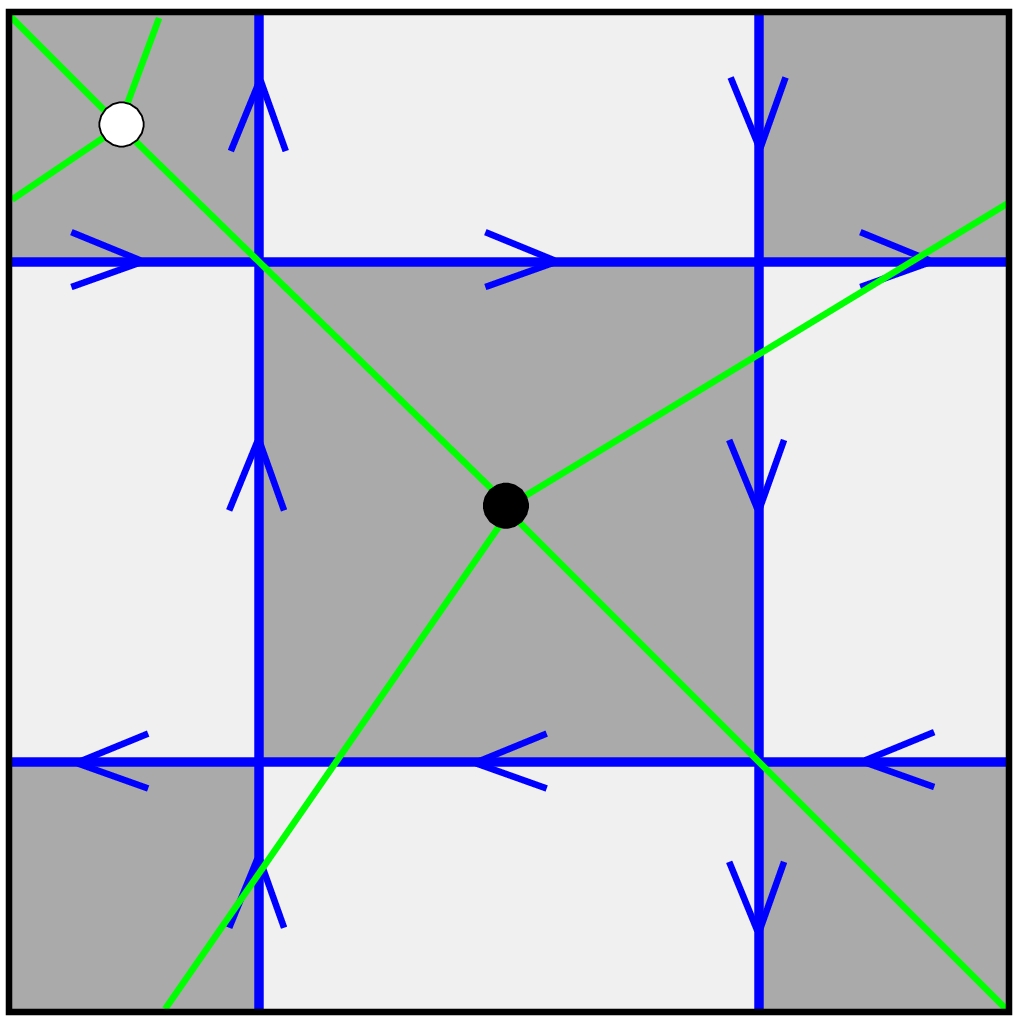}}
  \caption{Conifold rhombus loops and brane tiling}
  \label{fiaconi1}
\end{figure}

We now turn to the conifold and will see how we can reproduce the well--known square lattice
brane tiling for this theory (Figure 4). The toric diagram (\fref{tdconifold})
has four legs, these cycles can be seen in the rhombus loop diagram in \fref{fiaconi1}.

\begin{figure}[ht]
  \epsfxsize = 8cm
  \centerline{\epsfbox{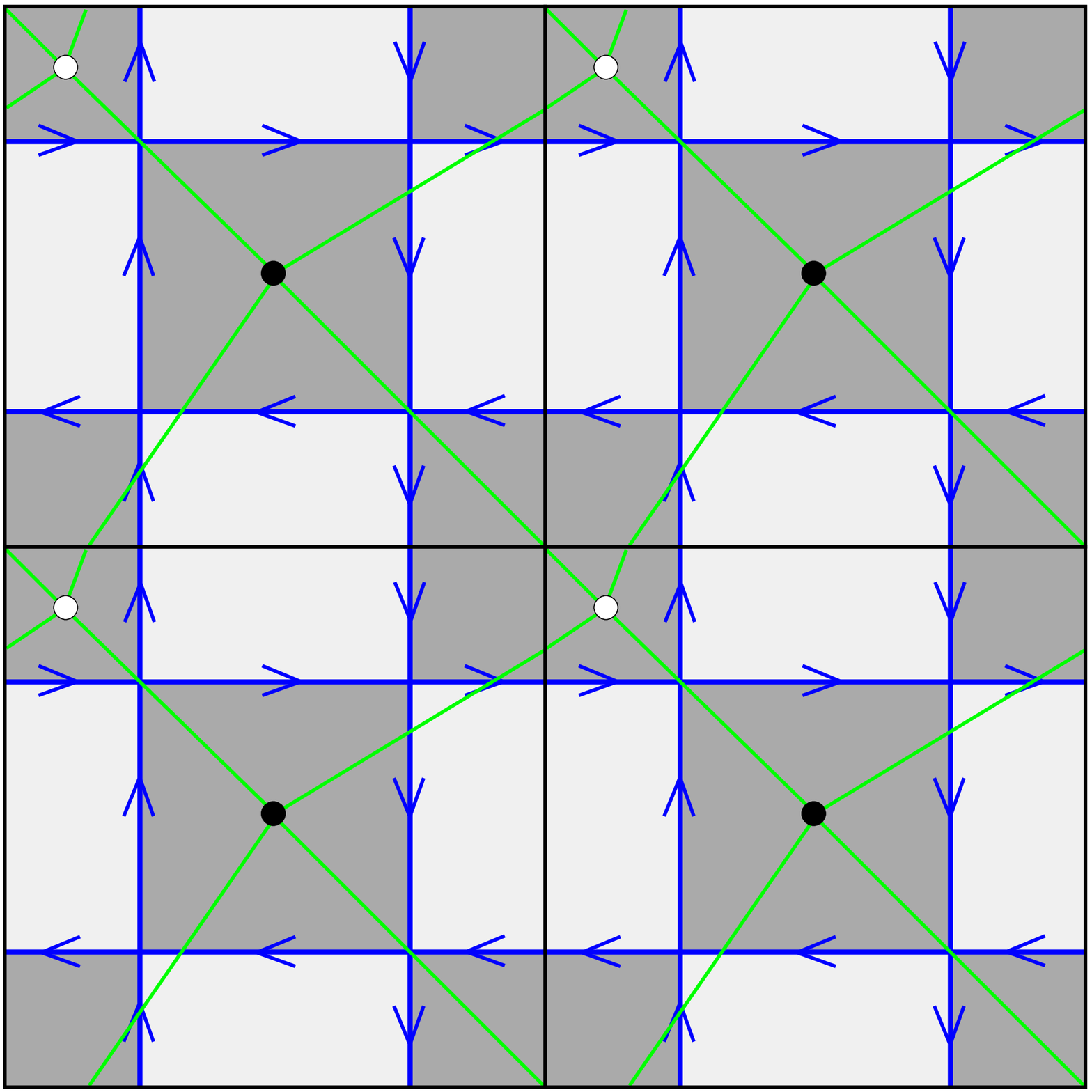}}
  \caption{Four fundamental cells of the conifold rhombus loop diagram.
If we consider these cells as one big fundamental cell then we gain the rhombus loop diagram
of the $\Z_2 \times \Z_2$ orbifold of the conifold.}
  \label{fiaconi2}
\end{figure}

We can actually skip the rhombus lattice step and draw the brane tiling immediately
in the rhombus loop diagram. The emerging green square tiling is better seen in \fref{fiaconi2}
where we have drawn a $2 \times 2$ block of adjacent fundamental cells. 
The square lattice tiling reproduces the superpotential of \cite{Klebanov:1998hh, Morrison:1998cs}.

\begin{figure}[ht]
  \epsfxsize = 9cm
  \centerline{\epsfbox{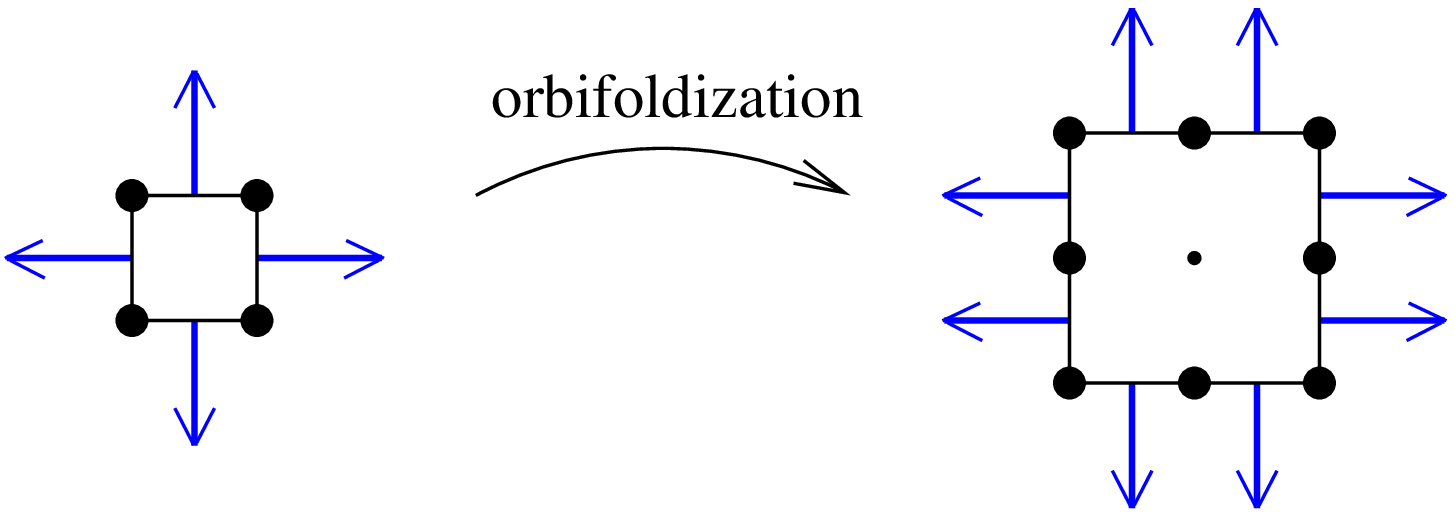}}
  \caption{$\Z_2 \times \Z_2$ orbifold of the conifold.}
  \label{tdconiorbi}
\end{figure}

Refining the integer lattice of the toric diagram means orbifoldizing the singular Calabi--Yau manifold.
The resulting toric diagram has more $(p,q)$--legs as seen in \fref{tdconiorbi}.
Clearly, we can realize orbifolding by increasing the size of the fundamental cell 
of the rhombus loop diagram to $n \times m$ times the size of the original cell.
This means orbifolding the space by $\Z_n \times \Z_m$. The action is generated by
\begin{eqnarray}
(z_1, z_2, z_3) \mapsto (\lambda z_1, z_2, \lambda^{-1} z_3),\hspace{0.2in} \lambda^m = 1 \\ 
(z_1, z_2, z_3) \mapsto (z_1, \omega z_2, \omega^{-1} z_3),\hspace{0.2in} \omega^n = 1
\end{eqnarray}

Multiplying the unit cell of the rhombus loop diagram is the same as 
increasing the size of the fundamental cell in the brane tiling
therefore it justifies the observations made in \cite{Hanany:2005ve}.

\subsection{$L^{131}$}

\begin{figure}[ht]
  \epsfxsize = 2.5cm
  \centerline{\epsfbox{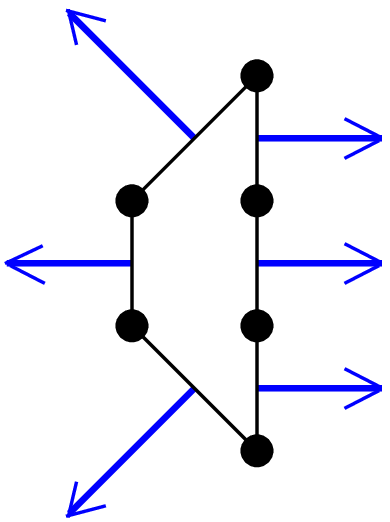}}
  \caption{$L^{131}$ toric diagram}
  \label{tdL131}
\end{figure}

\begin{figure}[ht]
  \epsfxsize = 6cm
  \centerline{\epsfbox{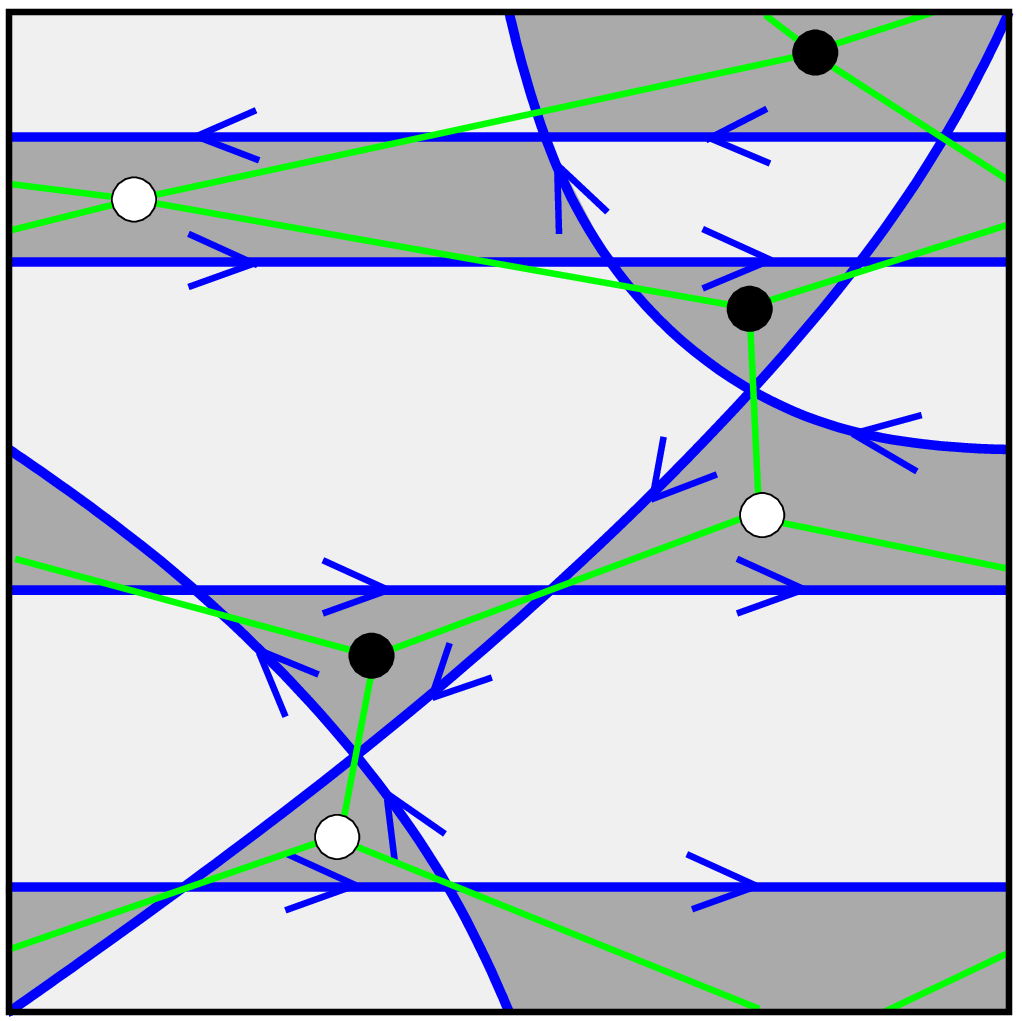}}
  \caption{$L^{131}$ rhombus loops and brane tiling}
  \label{fiaL131}
\end{figure}

As a more complex example, we generate brane tiling for $L^{131}$
which denotes one of the recently discovered 5d Sasaki--Einstein metrics 
(\cite{Cvetic:2005ft}, see also \cite{Cvetic:2005vk}). 
The space is topologically $S^2\times S^3$.
The toric diagram (\fref{tdL131}) has six legs, 
one possible rhombus loop diagram for them is shown in \fref{fiaL131}. We notice that by moving
the blue loops around, we may get a different tiling. This important phenomenon is toric duality
and will be investigated in section \ref{section_dualities}.

\begin{figure}[ht]
  \epsfxsize = 14cm
  \centerline{\epsfbox{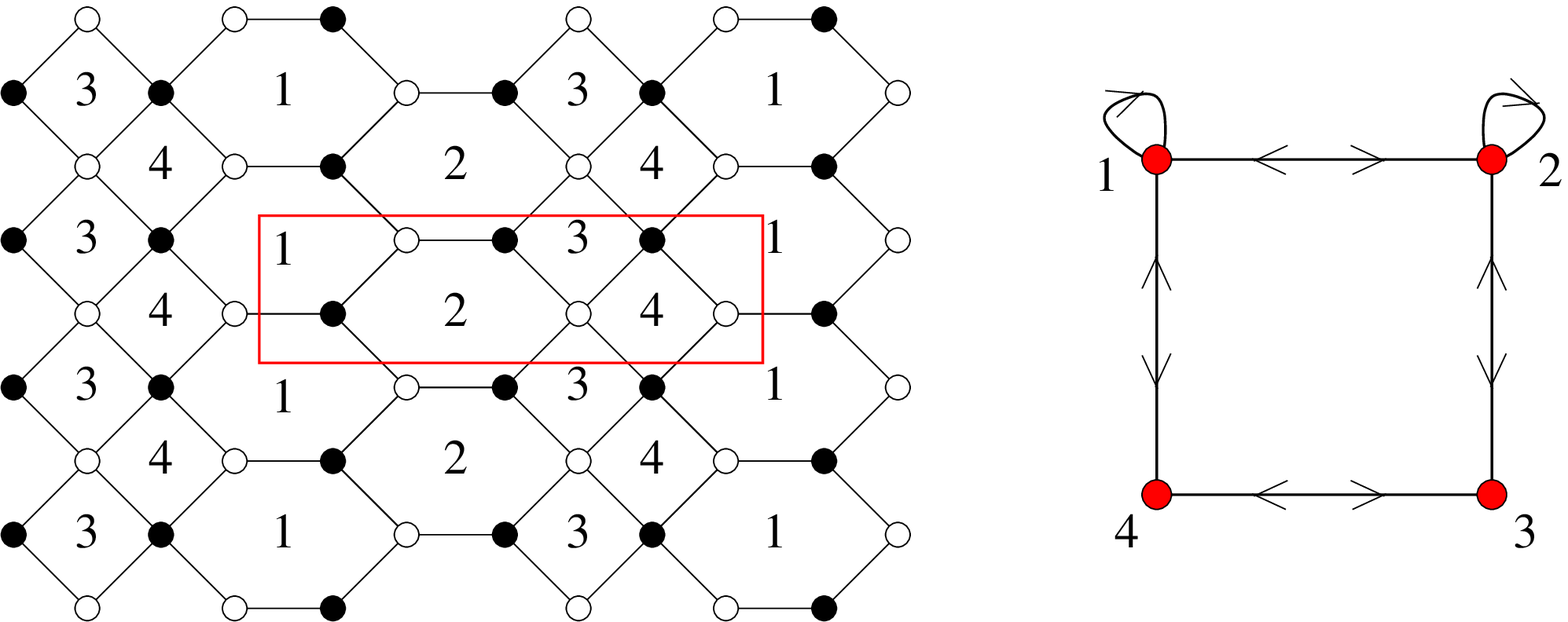}}
  \caption{(i) $L^{131}$ brane tiling (ii) and the corresponding quiver.}
  \label{L131tq}
\end{figure}

We can immediately draw the green tiling edges in the rhombus loop diagram, 
the final brane tiling can be seen in \fref{L131tq} (i). From the tiling we trivially obtain
the quiver (the ``compactified'' dual graph to the tiling, \fref{L131tq} (ii)) and the following superpotential:
\bea
  W=X_{11} X_{12} X_{21} + X_{22} X_{23} X_{32} + X_{43} X_{34} X_{41} X_{14} \\ 
 -  X_{21} X_{12} X_{22} - X_{32} X_{23} X_{34} X_{43} - X_{11} X_{14} X_{41} 
\eea

The fundamental cell in the tiling is denoted by a red box, this is the same as the fundamental cell
of the rhombus loop diagram.

Closed oriented loops in the rhombus loop diagram (\fref{fiaL131}) have a corresponding
{\bf gauge invariant trace operator} which is the product of the bifundamentals 
(at the intersection points) along the loop. These operators give a subset of all possible
gauge invariant operators.
Superpotential terms are trivial examples, these are small loops around the dark faces
in the diagram. Another example is provided by the zig--zag operator, for which the above mentioned 
oriented loop is just one of the rhombus loops.

If there are no degenerate rhombi, then we can use the results of \cite{Kenyon:rhombic}
and count the {\bf number of bifundamental fields} directly from the toric diagram.
This can be done by summing up the intersection numbers as in \cite{Hanany:2001py}. 
The number of fields coming from the crossing $(p_1,q_1)$ and $(p_2,q_2)$ rhombus loops is simply
\be
  \#(S_i \cdot S_j) = \#(C_i \cdot C_j) = |p_1 q_2-p_2 q_1|
\ee
Hirzebruch zero has two phases, one of them is non--degenerate (i.~e. there are no degenerate rhombi).
The formula gives the right value (eight) for the number of fields.
The other phase has $R_i=1$ for some of the bifundamentals, so the above formula
can't be used.

\subsection{$L^{152}$}

\begin{figure}[ht]
  \epsfxsize = 3.0cm
  \centerline{\epsfbox{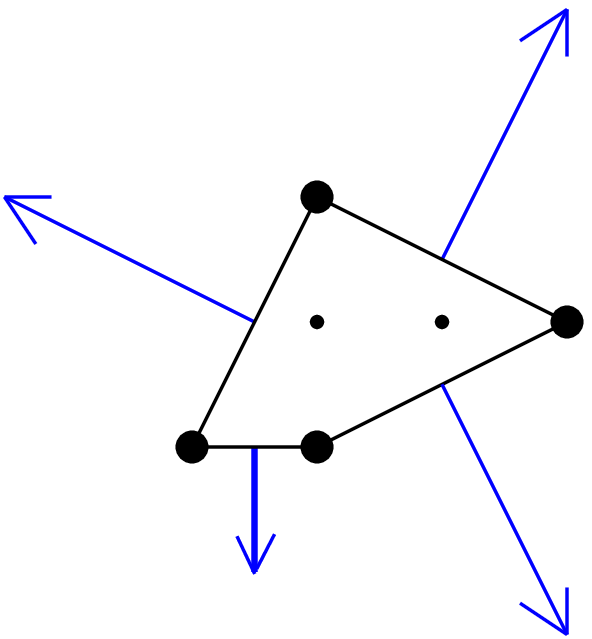}}
  \caption{Toric diagram of $L^{152}$}
  \label{tdnew}
\end{figure}

Our last example\footnote{
The quiver gauge theory for $L^{abc}$  has been constructed recently in 
\cite{newVegh, Benvenuti:2005ja, Butti:2005sw}.}  is $L^{152}$, its toric diagram is in \fref{tdnew}.

\begin{figure}[ht]
  \epsfxsize = 6cm
  \centerline{\epsfbox{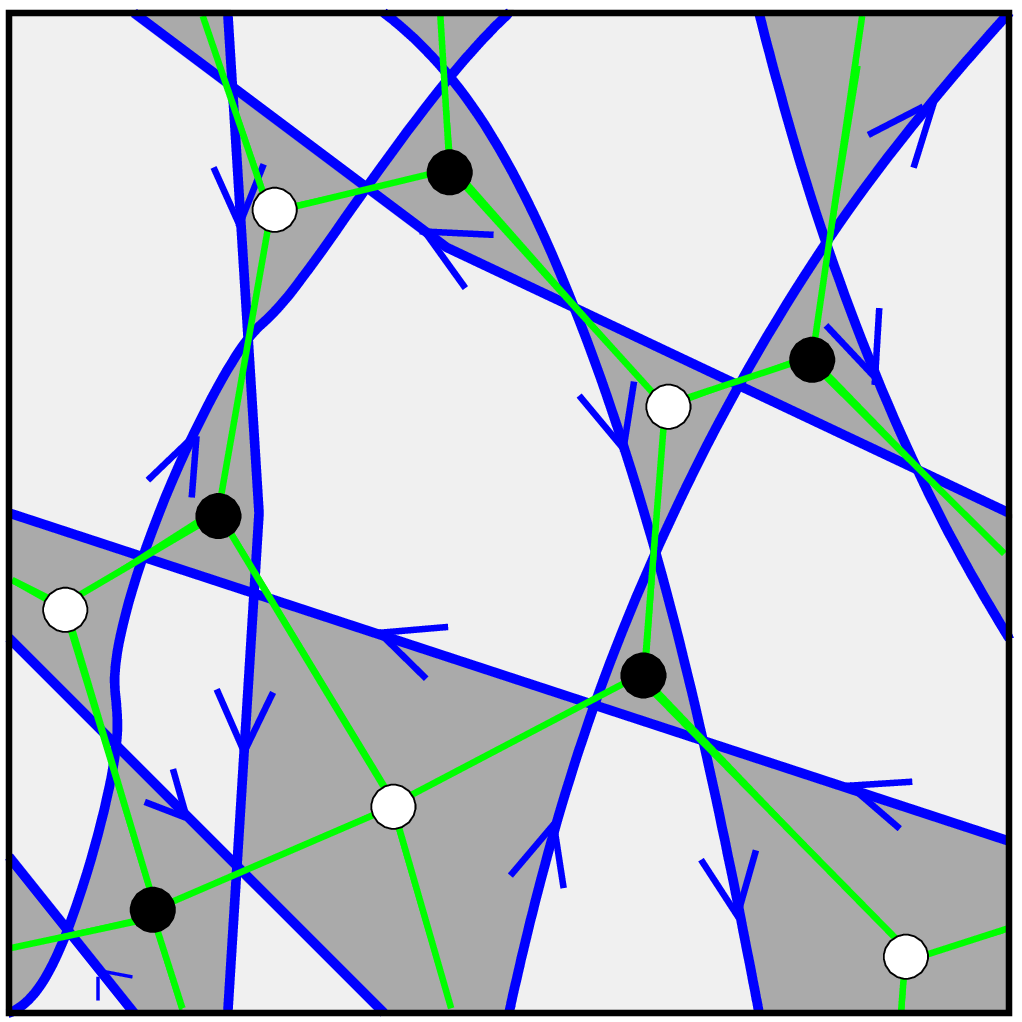}}
  \caption{$L^{152}$ brane tiling from the rhombus loops}
  \label{fiaNEW_4}
\end{figure}

The drawing of the rhombus loop diagram (\fref{fiaNEW_4}) is more involved than in the previous cases.
To obtain an anomaly free tiling, one has to make sure that every other face (the light areas)
has an even number of bounding rhombus loops (i.~e. in the tiling the the corresponding face
has even number of edges). To decide which face is dark and which one is light, we recall that the 
dark superpotential faces are distinguished by the fact that the rhombus
loops are oriented around them. The gauge invariant trace operators built up from these small oriented
loops are present in the superpotential, the order of the operator is given by the number of bounding 
rhombus loops of the dark face (this can be arbitrary).

\begin{figure}[ht]
  \epsfxsize = 9cm
  \centerline{\epsfbox{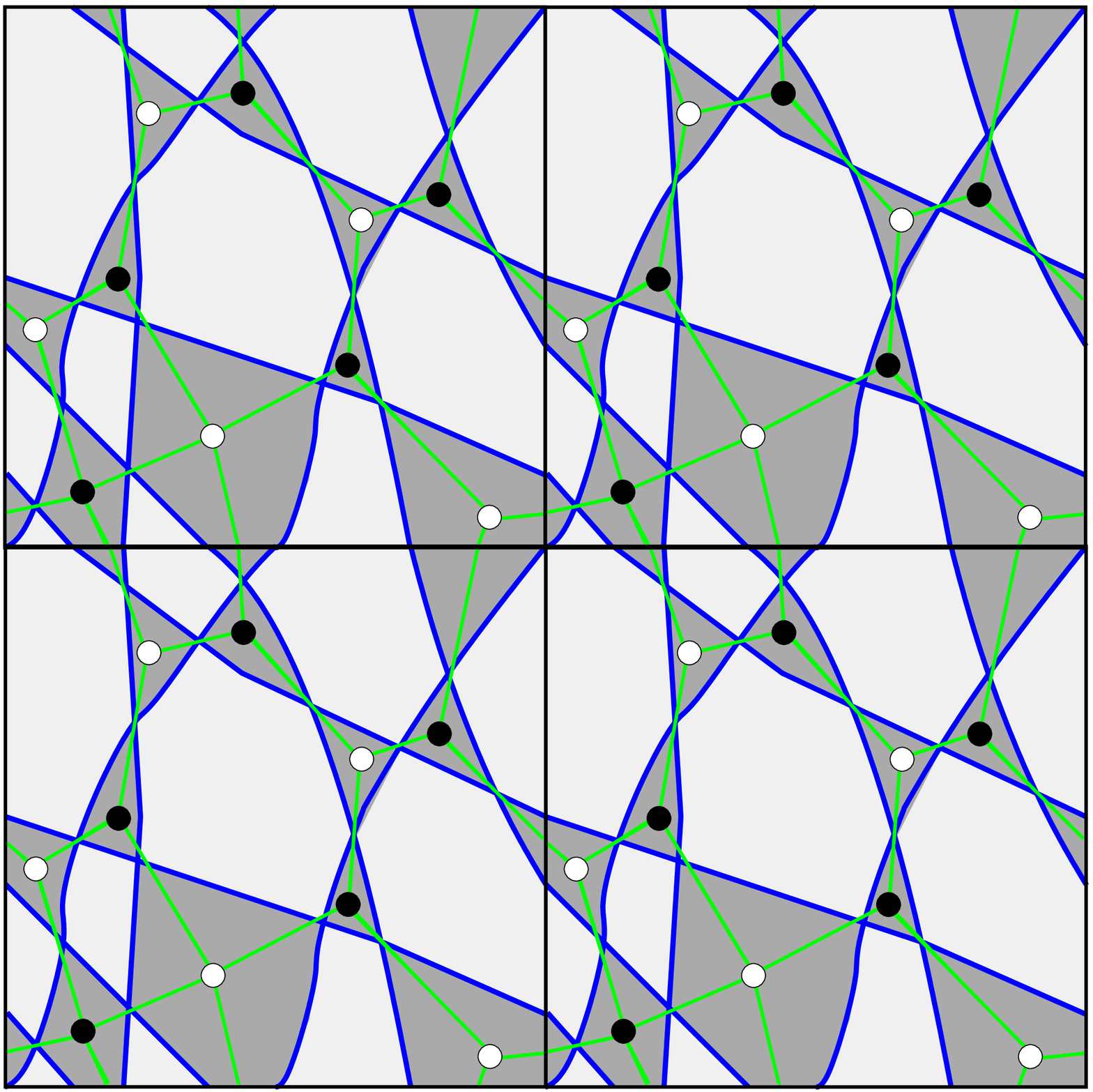}}
  \caption{$2 \times 2$ fundamental cells of the rhombus loop diagram of $L^{152}$. 
The brane tiling is shown in green.}
  \label{fiaNEWmore}
\end{figure}

Again, to see the tiling emerging out of the rhombus loop diagram, we have drawn more fundamental
cells next to each other (\fref{fiaNEWmore}). The dark faces get black and white nodes, 
the edges of the tiling are stretching between them.

\begin{figure}[ht]
  \epsfxsize = 14cm
  \centerline{\epsfbox{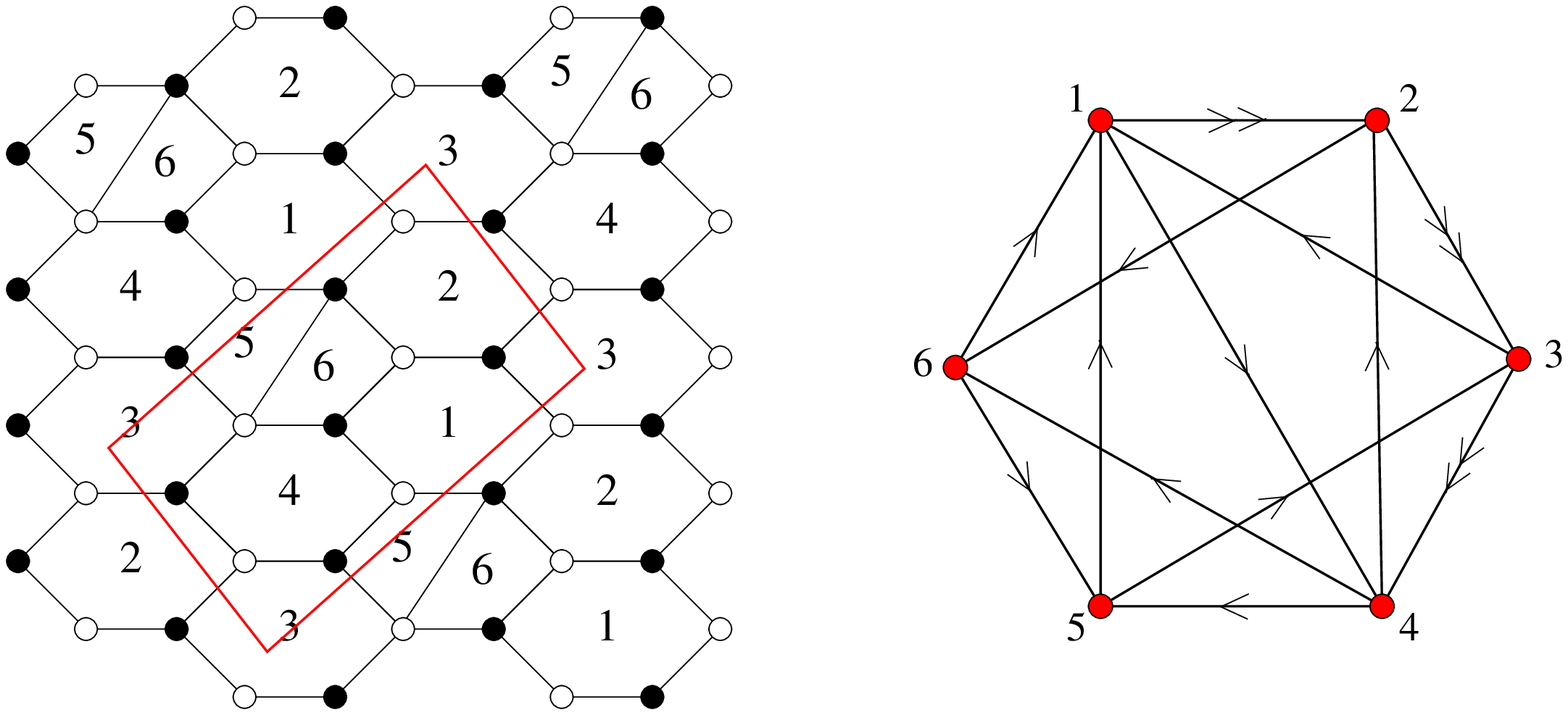}}
  \caption{(i) $L^{152}$ brane tiling (ii) and the corresponding quiver}
  \label{NEWtq}
\end{figure}

Finally, \fref{NEWtq} shows the resulting brane tiling and quiver.
Six gauge groups are present in the theory. This is what we expected from the area of the toric
diagram, as follows from equation \eqref{Pick}.

We can check the resulting tiling by computing the characteristic polynomial (\ref{P_L152}) of the dimer model
by means of the determinant of the Kasteleyn matrix (\ref{K_L152}) (for details see \cite{Franco:2005rj}).
The Newton polygon reproduces our starting point, the toric diagram of $L^{152}$ (\fref{tdnew})
therefore justifies our computation.

\beq
  K = \left( 
  \begin{array}{ccccc}  
    1 & 1 & -1 & w^{-1} & 0 \\
    w & 1 & 0 & 0 & z \\
    0 & 1 & 1 & 1 & 0 \\
    0 & 0 & w & 1 & 1 \\
    z^{-1} & 0 & 1 & 0 & 1 
  \end{array} 
  \right)
  \label{K_L152}
\eeq

\be
  P(w,z) \equiv \mbox{det}(K) = 6-6w+w^2+z^{-1}+w^{-1}z^{-1}+z
  \label{P_L152}
\ee

\section{Toric duality and Seiberg duality}
\label{section_dualities}

We have seen ambiguities while constructing the brane tilings for a given singularity.
{\bf The non--uniqueness manifests itself through the fact that we can freely move the rhombus loops}
which certainly changes the tiling and therefore the quiver gauge theory.
Some of the resulting tiling might not be bipartite. Out of the bipartite tilings we are also
only interested in the consistent ones.
These ``phases'' of the theory are believed to be Seiberg--dual to each other 
\cite{Beasley:2001zp, Feng:2001bn, Cachazo:2001sg, Berenstein:2002fi}.

\begin{figure}[ht]
  \epsfxsize = 9cm
  \centerline{\epsfbox{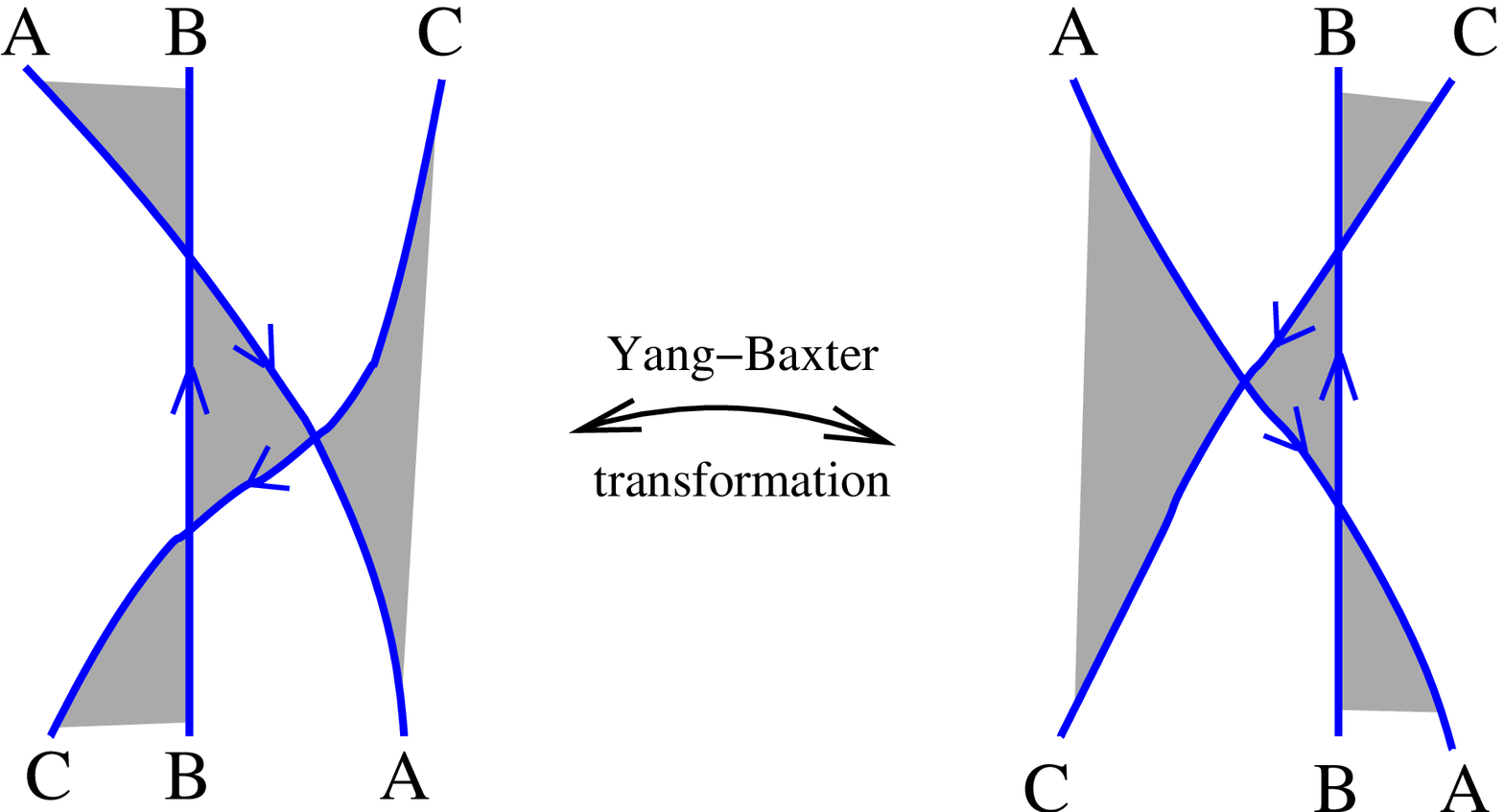}}
  \caption{The elementary Picard--Lefschetz--Yang--Baxter transformation.}
  \label{yangbaxter1}
\end{figure}

\begin{figure}[ht]
  \epsfxsize = 11cm
  \centerline{\epsfbox{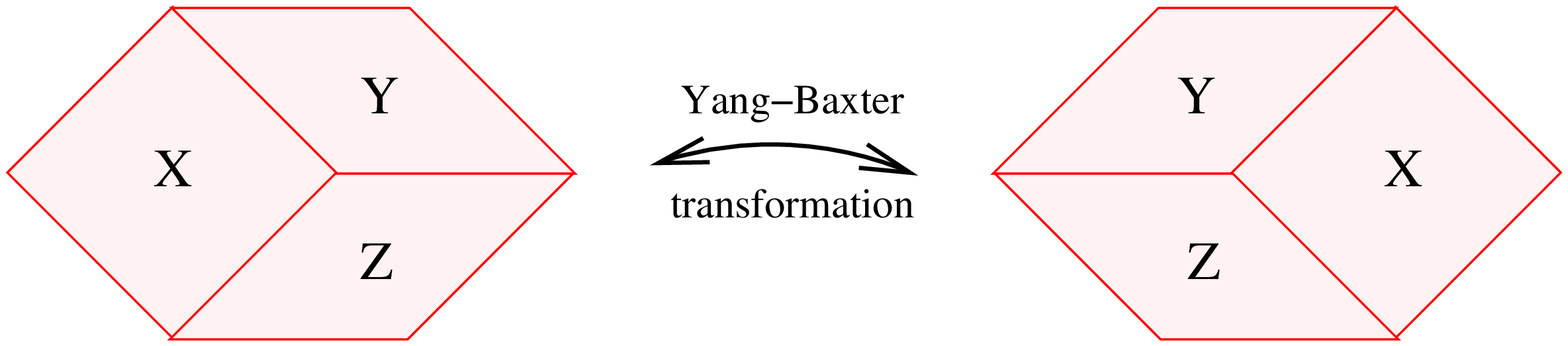}}
  \caption{The Yang--Baxter--Reidemeister transformation on the rhombus lattice.
Star-triangle}
  \label{yangbaxter2}
\end{figure}

The simplest transformation is when we move a single rhombus loop across an intersection point
as in \fref{yangbaxter1}. This is the {\bf Yang--Baxter transformation}. 
We can build up a generic transformation from such elementary steps.
The Yang--Baxter move changes the rhombus lattice locally which is shown in \fref{yangbaxter2}.

On the other hand, the brane tiling (and the periodic quiver) has been changed globally.
Apart from the local change in the rhombus lattice, we are forced to ``flip'' the
tiling edges in the rhombi (\fref{rhombus1}), i.~e. the periodic quiver and the brane tiling get interchanged.
The periodic quiver is usually non--bipartite (the only exception is the square lattice),
therefore the resulting tiling is non--bipartite. However, one can perform more such
Yang--Baxter transformations so that the final brane tiling is anomaly--free.
Then, by definition, the resulting theory is {\bf toric dual} to the original one.
We provide an example in the followings.

\subsection{Seiberg duality in the hexagonal lattice with extra line}

\begin{figure}[ht]
  \epsfxsize = 12cm
  \centerline{\epsfbox{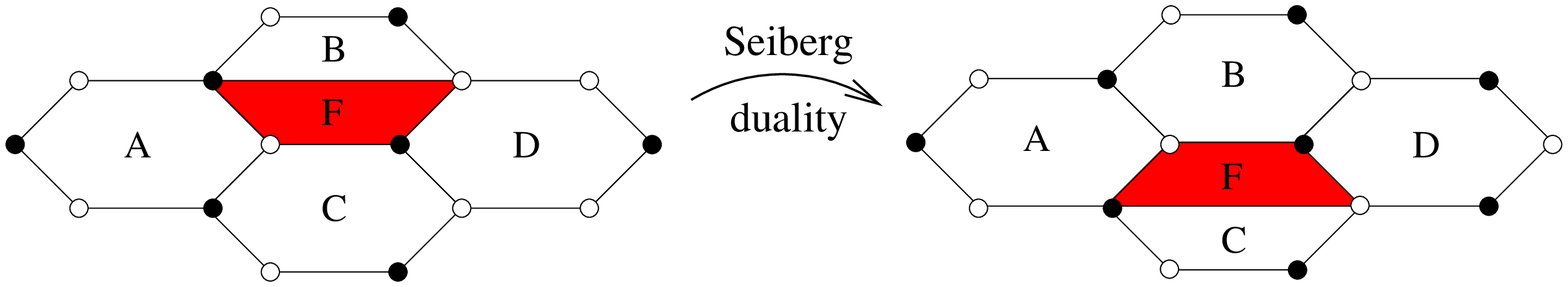}}
  \caption{(i) Four hexagon with one extra line. (ii) Seiberg dualizing the red square ($F$).
The extra edge in the upper hexagon ($B$ \& $F$) gets into the lower one ($F$ \& $C$).}
  \label{seiberg_hexagon}
\end{figure}

Let us consider an arbitrary brane tiling with a subtiling shown in \fref{seiberg_hexagon} (i).
This setup has been used in \cite{newVegh}. If we dualize group $F$, the extra edge moves
into the neighboring hexagon. 

\begin{figure}[ht]
  \epsfxsize = 17cm
  \centerline{\epsfbox{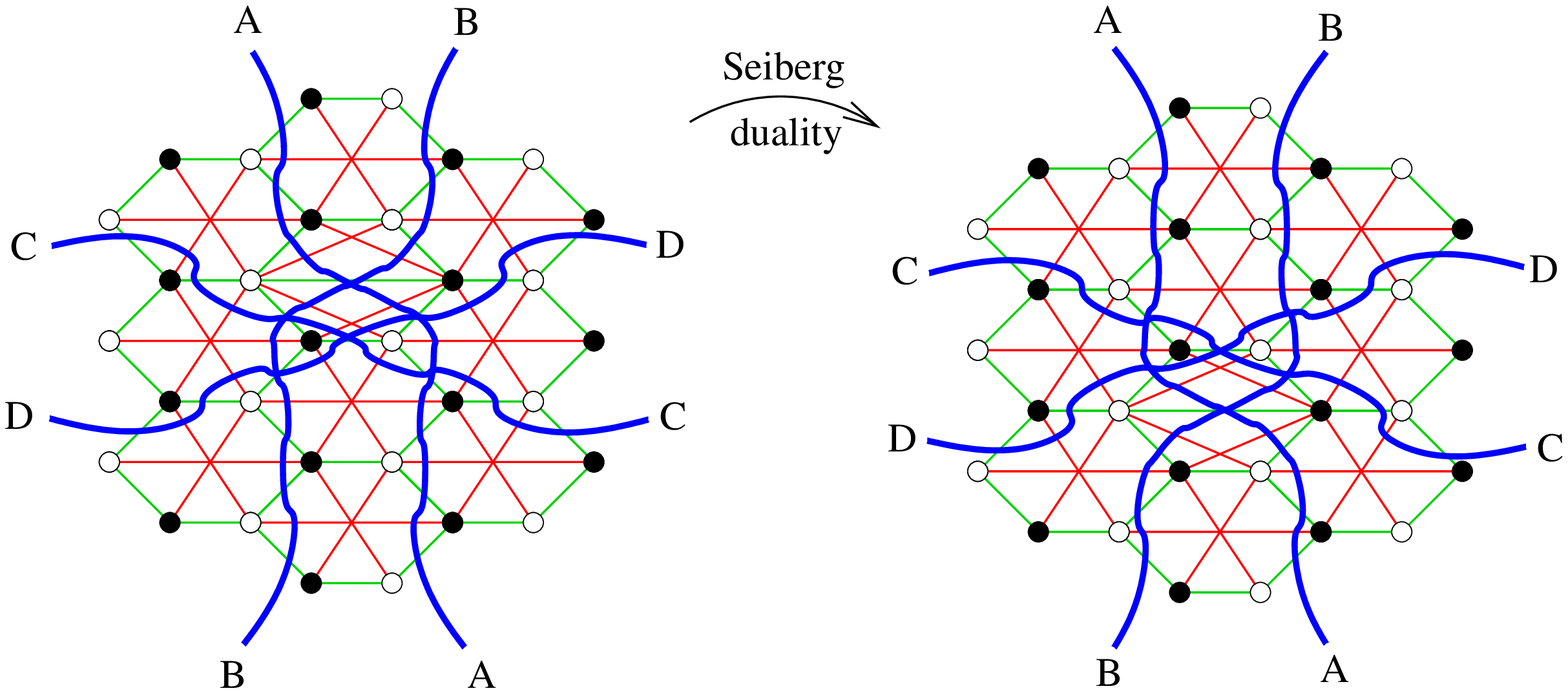}}
  \caption{Seiberg duality in the hexagonal tiling with extra edge. 
The brane tiling is shown in green, the (deformed) rhombus lattice is in red, the relevant rhombus
loops are in blue.}
  \label{seiberg_hexagon2}
\end{figure}

What happened to the rhombus loops during this dualization? 
We can see that immediately, if we draw the (red) rhombus lattice (\fref{seiberg_hexagon2}).
The relevant rhombus loops ($A$,$B$,$C$,$D$) are shown in blue as usual. Only these loops are affected
by the transformation.

\fref{seiberg_hexagon} shows the rhombus loops only. 
In this picture we see how Seiberg duality can be realized on the level of rhombus loops.
It can be easily checked that the transformation contains four elementary Yang--Baxter steps.
For another brane realization of Seiberg duality see \cite{Hanany:1996ie,Elitzur:1997fh}.

\begin{figure}[ht]
  \epsfxsize = 15cm
  \centerline{\epsfbox{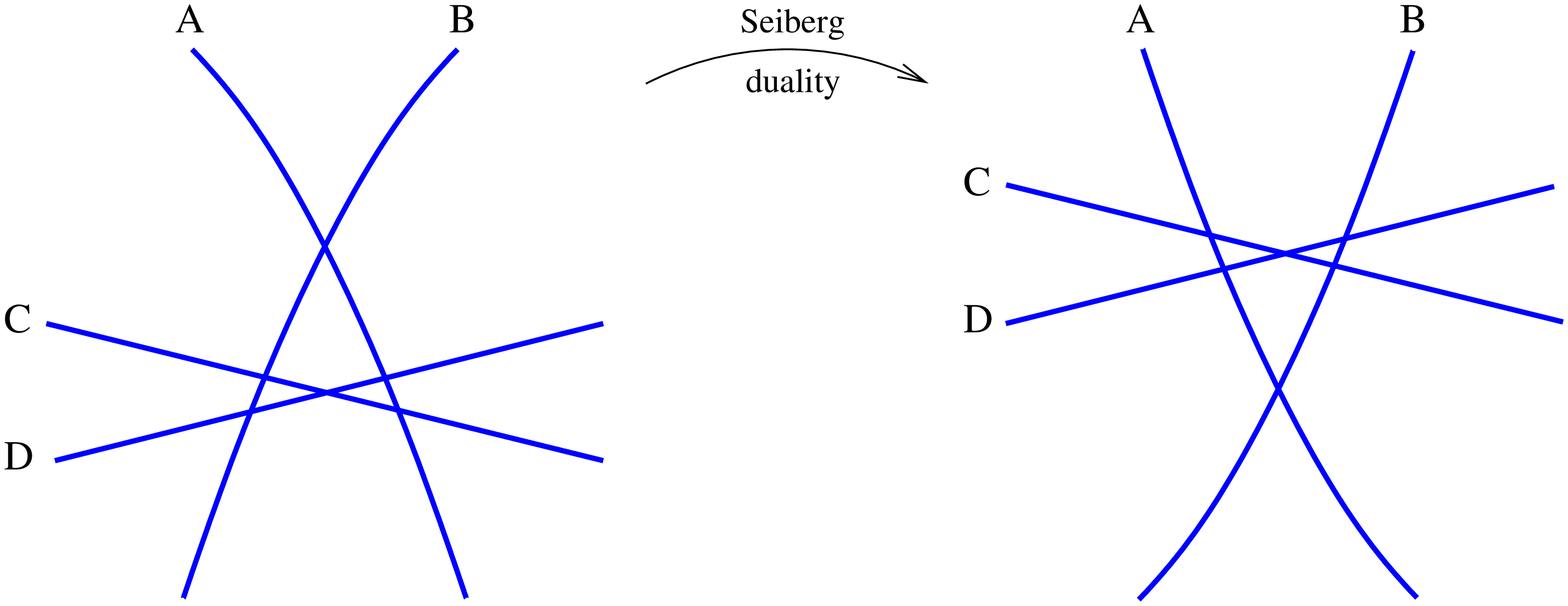}}
  \caption{Seiberg duality in the level of the rhombus loops.}
  \label{seiberg_hexagon3}
\end{figure}

With this knowledge, a thorough study of the possible moves of the rhombus loops 
(and the inconsistencies of the tiling) should reveal whether or not Seiberg duality 
is equivalent to toric duality.

\newpage
\section{Conclusions}
\label{section_conclusions}

In this paper we have proposed the {\bf Fast Inverse Algorithm} that computes the
quiver gauge theory living on probe D3--branes. The algorithm computes the
recently discovered {\bf brane tilings}, i.~e. the quiver gauge theory directly from the toric diagram
of the AdS/CFT dual singularity without using the metric.
It gives a better understanding of the connection of the quiver theory and the singularity.

In the following we summarize by points some open questions and potential directions for future research in the hope that some of these ideas will get more concrete realization.

In \cite{Baxter:1986df} Baxter introduced the so--called {\bf Z--invariant Ising model}.
The construction is based on
rhombus loops (a.k.a. rapidity lines, see Figure 7 in \cite{Baxter:1986df})
and the rhombus loop diagram. 
This is the most general setting in which the Ising model is exactly solvable.
It would be interesting to study this model from the viewpoint of string theory
and integrable structures in $AdS \times X$ spaces.

By simply drawing an arbitrary periodic bipartite graph we might get an inconsistent theory.
The {\bf full classification of consistent brane tilings} is still an unsolved problem, although
the rhombus lattice technique presented in this paper gives
partial solution by giving constraints on the consistent tilings.

As is well known, a given singular manifold may have many toric phases. The number of such toric phases may have an important role for the geometry. These phases differ from each other by the number of fields, the tiling configuration and by the number of terms in the superpotential. This therefore leads to a natural question. Is there a phase which is more spacial or more fundamental than the other phases? We may call this phase the {\bf canonical phase}. Can a {\bf canonical phase} be defined for each quiver theory?
A starting point for answering this can be the lemma of Kenyon, \cite{Kenyon:2003uj}, which states that the hexagonal lattice is
universal, i.~e. every tiling can be obtained by removing edges from the hexagonal
lattice (and integrating out two--valence nodes). This statement is also true for the square lattice.
In all examples studied so far it appears that the squares and the 
hexagons are the basic building blocks for at least one representative 
from the set of toric phases of a given toric singularity. All toric 
models seem to interpolate between hexagons and squares. It will be interesting to find more evidence for this observation.

The brane tiling is periodic, or equivalently, it is sitting on a 2--torus.
An interesting generalization would be the extension of the idea
to higher genus surfaces. This might be done in the context of {\bf discrete Riemann
surfaces} \cite{mercat}.

We have seen that Seiberg duality can be realized by moving around the rhombus loops.
Nevertheless, not all deformations of the rhombus loops give an anomaly--free
quiver (i.~e. bipartite tiling), and even the anomaly--free ones can be tachyonic --
some of the R--charges may vanish.
It would be intriguing to understand which brane tilings are ``good'' in the above
mentioned sense, this might shed some light on how to {\bf generalize
Seiberg duality}.

Another interesting direction is to study the Fast Forward and Fast Inverse Algorithms in the context 
of the {\bf derived category} approach to D--branes on Calabi--Yau manifolds \cite{Aspinwall:2004jr, newHHV}. One can also study the direct {\bf equivalence of the Forward and the Fast Forward Algorithms}
\cite{newFV}. This gives another justification for brane tilings.
Recent results in understanding the Fast Inverse Algorithm in an {\bf intersecting D6--brane}
scenario will be presented in \cite{newFHKV}.

\vskip 0.5cm

{\bf Acknowledgements:} 
We gratefully acknowledge the invaluable discussions we have had with
S.~Benvenuti, S.~Franco, C.~Herzog, P.~Kazakopoulos, K.~D.~Kennaway, A.~King, 
T.~Okuda, S.~Pinansky and Y.~Tachikawa. 
We are indebted to B.~Feng and B.~Wecht for reading the draft and making many comments.
We thank the KITP for hospitality during part of this work.


\newpage

\bibliography{paper}
\bibliographystyle{JHEP}

\end{document}